\documentclass[prd,showpacs,amsmath,amssymb,nobibnotes,nofootinbib,11pt]{revtex4-1}
\usepackage{appendix}
\usepackage{hyperref}
\usepackage{graphicx}
\usepackage{algorithm}
\usepackage{algorithmic}
\usepackage{amsmath}
\usepackage{txfonts}
\usepackage{subfigure}
\usepackage{amssymb}
\usepackage{subfigure}
\usepackage{multirow}
\usepackage{youngtab}
\usepackage{colortbl}
\usepackage{slashbox}

\usepackage{hhline}
\newtheorem{theorem}{Theorem}[section]
\newtheorem{definition}[theorem]{Definition}

\newenvironment{defn*}{\begin{definition}}{\end{definition}}

\def\be{\begin{equation}} \def\ee{\end{equation}} \def\ua{\dot{x}}
\def\bea{\begin{eqnarray}} \def\eea{\end{eqnarray}}
\def\fii{\hat{\phi}^{(i)}}

\def\mi{m^{(i)}}\def\mk{m^{(k)}}

\def\tkl{{t^{\kappa ...}}_{\lambda ...}}

\def\emt{T^{\alpha \beta}}
\def\bs{\begin{split}}
\def\es{\end{split}}
\def\df{\dfrac}

\def\skl{\hat{T}^{\kappa \lambda}(z(s), X)}
\def\keg{\sqrt{-g}}

\begin{document}
\title{Dynamics of Extended Objects in General Relativity}

\author{\.{I}brahim  Burak \.{I}lhan}
\email{iburak@metu.edu.tr}
\affiliation{Department of Physics, Middle East Technical University,
Ankara, Turkey}

\begin{abstract}
In this thesis, multipole expansions of mass, momentum and stress density will be made for a body in Newtonian mechanics. Using these definitions; momentum, angular momentum, center of mass, force and torque are defined for $N$ gravitationally interacting isolated bodies. Equations of motions of such a system are derived. Definitions of momentum, angular momentum, center of mass, force and torque are made in a relativistic theory. Dynamical (gravitational) skeleton is defined and the multipole moments of the dynamical skeleton are found. Equations of motion for a test body moving in a gravitational field are derived in terms of the multipole moments. Save the details of the derivations, no originality in this thesis is claimed: it is intended as a review of the subject.
\end{abstract}

\maketitle
\tableofcontents
\newpage

\section{INTRODUCTION}
\label{sec:introduction}

Describing the motion of an object in a gravitational field is the main concern of any theory of gravity. But even in the simplest cases, the problem is a difficult one when one aims for an accurate description of Nature. Dealing with the motion of a point particle is a simple one, but when some structure is added to the object in concern, the problem gets complicated. A solution for the problem of motion of $N$ isolated bodies with internal structure in Newtonian mechanics was first proposed by Tisserand \cite{tiss}. The method suggested a way for separation of the external motion of the bodies from their internal motion, by making use of the linearity of equations of motion. Reviews of this method can be found in \cite{damour}. As a result of this separation, overall motion of one of the bodies can be described to the order of desired accuracy, and justification for the point particle approximation can be made more reliable \cite{dixonIV}.

In General Relativity, the field equations are coupled and nonlinear, and these prevent direct application of the above mentioned methods of Newtonian mechanics. Even solving the equation of motion for a point particle is not straightforward, since the existence of a particle changes the metric tensor. If this effect (self-field) is neglected, it was shown in \cite{schild} that such a particle follows a geodesic. Introduction of the self-field highly complicates the problem \cite{poisson, waldself}. Ignoring the self force, a way of approaching the problem of motion of an extended test body in General Relativity was proposed by Mathisson in 1937 \cite{mat}. The idea was to define the body by a set of multipole moments defined on a central worldline, which is to be identified as the generalization of the Newtonian concept of mass center to a relativistic theory. His paper showed that in a relativistic theory one can have the useful definitions of  mass center, force, torque and mass; with desired properties. Mathisson called the set of moments as the ``Dynamical (or Gravitational) Skeleton", and the resulting equations of motion as the ``Variational Equation of Dynamics". A similar treatment of the problem was proposed by Papapetrou \cite{papa}, and Mathisson`s methods were further developed by Tulczyjew \cite{Tul}, Tulczyjew \& Tulczyjew \cite{tul1} and others, with the accompanying improvements of the definition of mass center by Moller \cite{Tul, mol, bb} and others \cite{damour, dixonrev}.
No claim of originality is made in this thesis, except for the details of derivations. I mainly follow the series of papers Dixon wrote in this topic, expounding on these derivations \cite{dixonIV, dixon64, dixonI, dixonIII,  dixonV}.

In the first chapter, I will review the problem of motion of $N$ extended isolated bodies in Newtonian theory, and try to obtain some insight which will be useful in the relativistic theory. The first section will be devoted to the definitions of multipole moments in Newtonian theory. Moments of mass, momentum and stress, and their interrelations will be defined. In the second section, ``Reduced Moments" will be defined with the help of the previously defined moments, and the set of equations they satisfy will be found. In the third section, the simplest approximation to such a system - point particle approximation - will be made. This approximation ignores any kind of internal structure present in the bodies. In this case, only the masses of the bodies and their mutual separation play a role in the equations of motion. When some structure is added to the objects (quadrupole approximation), pole-dipole interaction will be observed. Considering higher order moments will give more and more accurate results, but these results are indeterminate, as will be seen, without an accompanying equation of state. To get determinate equations, one can work with rigid objects. This discussion will finalize the first chapter. The first chapter is based on \cite{damour, dixonIV, dixonV}.

In the second chapter, the analogous relativistic problem will be reviewed. At first, the difficulties faced during the generalization of the ideas from Newtonian mechanics will be discussed. The following three sections will discuss the definitions of momentum, angular momentum, center of mass worldline in General Relativity. The fifth section involves the discussion of the energy momentum skeleton, and sketches the outline of the existence proof of such an object. Also, its relation to the energy momentum tensor will be given explicitly. The rest of the chapter involves the definition of gravitational force and torque, and the multipole moments of the energy momentum skeleton. This chapter will mainly review \cite{dixonIV, dixonI, dixonIII, de felice, schattnerdixon}.

During the development of the theory, several mathematical tools are needed. Two-point tensors - especially Synge's world function - are frequently used in the relativistic theory, and discussed in detail in Appendix \ref{sec:world function}. Useful definitions of vertical and horizontal covariant derivatives, defined for a two-point tensor, is discussed in Appendix \ref{sec:Vertical and Horizontal Covariant Derivatives}. Jacobi equation and its consequences are frequently used, and is discussed in Appendix \ref{sec:Jacobi Equation}.

Moreover, tensor extensions, which are used in the definitions of relativistic force and torque, is important, and its definition, in addition to normal coordinates, can be found in Appendix \ref{sec:Tensor Extension}
. The appendices are mainly based on \cite{synge, dixonIV, schattnerjacobi, veblen}, respectively.

\subsection{NOTATION AND CONVENTIONS}
\label{sec:notation}
Throughout this thesis, Greek indices will run from 0 to 3 and Latin ones from 1 to 3. The metric is of the signature $(+,-,-,-)$. Derivation along a curve $x(u)$ is denoted by $\df{\delta}{\delta u} \equiv \df{d x^\alpha}{du} \nabla_\alpha$. Riemann tensor is defined through the relation
\be [\nabla_\alpha,\nabla_\beta] \xi^\gamma = {R_{\alpha \beta \delta}}^\gamma \xi^\delta.\ee
Two-point tensors are frequently used, and a special notation is introduced for notational simplicity. When dealing with a two-point tensor, $\alpha,\beta , \gamma,...$ will be used as tensor indices at $x$, and $\kappa,\lambda , \mu,...$ for tensor indices at $z$; unless stated otherwise.  Covariant derivatives of the world function will be denoted by subscripts, such that
\be \nabla_\alpha \nabla_\kappa \sigma(x,z)= \sigma_{\kappa \alpha}(x,z). \ee
Curly bracket notation, defined only for 3 indices, is defined as:
\be t_{\{\alpha \beta \gamma\}} \equiv t_{\alpha \beta \gamma} - t_{\beta \gamma \alpha} + t_{\gamma \alpha \beta}.\ee
Throughout the thesis, round brackets will denote symmetrization, such as:
\be t_{(\alpha \beta)} \equiv \df{1}{2}(t_{\alpha \beta} + t_{\beta \alpha}), \ee
and square brackets will denote antisymmetrization:
\be t_{[\alpha \beta]} \equiv \df{1}{2}(t_{\alpha \beta} - t_{\beta \alpha}). \ee

\section{NEWTONIAN MECHANICS}
\label{sec:Newtonian Mechanics}

In this chapter, we will deal with the problem of motion of $N$ isolated extended bodies in detail, in Newtonian mechanics. This will give us an idea about the necessary steps to be taken in a relativistic theory. Our main aim will be the  separation of the external motion of the bodies from their internal motion.

Consider a system composed of $N$ bodies such that the individual size of each body is small compared to the mutual separation between each of them; and let each body be made of some fluid, with an internal structure governed by a given equation of state.

Let $\vec{x}(t)$ be the position vector of a general point with respect to a fixed origin and $\rho (\vec{x},t)$ be the mass density. (In the following lines, index notation, such as $x_a$, will be used instead of vector notation for simplicity.) Save the details of the derivations, nothing we shall present is new. We follow \cite{damour, dixonIV, landaufluid}.

The equations governing the motion are:
\begin{enumerate}
  \item Mass Conservation - Continuity Equation:
  \begin{equation}\dfrac{\partial \rho}{\partial t} + \dfrac{\partial}{\partial
  x^a} (\rho \dot{x}_a) = 0, \label{mc}
  \end{equation}
  which relates the rate of change of mass in some volume $V$ to the mass of fluid flowing through the boundary of $V$, and,
  \item Continuum Equation of Motion:
  \begin{equation}\dfrac{\partial (\rho \dot{x}_a)}{\partial t} +
  \dfrac{\partial}{\partial x^b}\bigg(\rho
  \dot{x}_a\dot{x}_b - \sigma_{ab} \bigg) = \rho \dfrac{\partial}{\partial x^a} \phi ,
  \label{cem} \end{equation}
  which is obtained by evaluating the rate of change of momentum density, by using Euler's equation that tells us that some fluid enclosed by a volume element exerts a pressure on that element.
\end{enumerate}
Here $\dot{x}_a \equiv \dfrac{dx_a}{dt}$, $\sigma_{ab}$ is the internal stress tensor - it contains the information of the pressure and possibly viscosity - and $\phi$ is the gravitational field given by the Poisson equation:

\be
{\nabla}^2\phi = -4 \pi G \rho , \label{poisson}
\ee
with the boundary condition $$\lim_{r \to \infty} \phi = 0, \quad \mbox{where} \quad r^2 \equiv x^a x_a. $$

Since the equations are coupled and nonlinear, instead of directly trying to solve them, one can try to find approximate solutions. The bodies are largely separated, so in order to solve the equation of motion of each body, one can make multipole expansion of the necessary function with respect to some origin, and keep terms up to the order of needed accuracy. The simplest case is to keep only the lowest order terms in each of the resulting series. This will give the point particle approximation for each body. Now, this procedure will be discussed in detail, in the following order \cite{damour, dixonIV}:
\begin{itemize}
  \item Multipole Expansion
  \item Reduced Moments
  \item Point Particle Approximation
  \item Internal Structure and Higher Moments
  \end{itemize}
\subsection{Multipole Expansion}
\subsubsection{Moments of a Scalar Function in $\mathbb{E}^3$}
\label{subsubsec:moments}Before dealing with the actual problem, we will first consider the multipole structure of a scalar function and see how the moments determine it completely. For this, consider a continuous scalar function $f(\vec{x})$ of compact support on $\mathbb{E}^3$. We ask for compact support since functions representing a finite sized object (mass, momentum or stress density) will be of this kind. Moments of such a function with respect to origin is defined as

\be F^{a_1...a_n} \equiv \int x^{a_1}...x^{a_n} f(\vec{x})d^3\vec{x}, \quad n \geq 0.\label{momentler}\ee
One can show that this set of moments completely describe the original function $f(\vec{x})$ by considering the Fourier transform of it, defined as

\be \tilde{f} (\vec{k}) \equiv \int f(\vec{x}) exp (i \vec{k} \cdot \vec{x})d^3\vec{x}.\ee
Expanding the exponential, this can be written in terms of (\ref{momentler})

\be \tilde{f}(\vec{k}) = \sum_{n=0}^\infty \int f(\vec{x}) \df{i^n}{n!} k_{a_1}... k_{a_n}x^{a_1}...x^{a_n} d^3\vec{x} = \sum_{n=0}^\infty  \df{i^n}{n!} k_{a_1}... k_{a_n}F^{a_1 ... a_n}. \label{ftild}\ee
And finally, one can obtain $f(\vec{x})$ by taking the inverse Fourier transform of this expression. $\tilde{f}(\vec{k})$ is called the ``modified moment generating function" of $f(\vec{x})$ \cite{dixonI}.
\subsubsection{Mass, Momentum and Stress Moments}
Now, consider one of the bodies in the system described at the beginning of this chapter. As a representative point for this particular body, introduce the moving origin $z_a(t)$, with respect to which the multipole expansion will be made. Let the velocity of this point be $\dfrac{dz_a} {dt} \equiv v_a$. Then, the position vector relative to this moving origin will be $r_a (t)= x_a (t) - z_a (t)$. We define moments of mass, momentum and stress densities with respect to this point, respectively, as

\be m_{a_1 a_2 ... a_n} \equiv \int r_{a_1} r_{a_2} ... r_{a_n} \rho d^3x \label{mm}, \ee

\be p_{a_1 a_2 ... a_n b}\equiv \int r_{a_1} r_{a_2} ... r_{a_n} \rho \dot{x}_b d^3x \label{pm}, \ee

\be t_{a_1 a_2 ... a_n b c} \equiv \int r_{a_1} r_{a_2} ... r_{a_n} (\rho \dot{x}_b \dot{x}_c - \sigma_{bc} ) d^3x \label{tm}, \ee
where the integrals are taken over some spatial volume containing the body that is being dealt with, but not any other body \cite{dixonIV, dixonV}.
 This can be done, since, at the beginning, it was assumed that the separation
 between the bodies is large. Note that for the sake of notational simplicity, we supress the arguments of the functions. (\ref{mc})
 and (\ref{cem}) relate time derivatives of mass and momentum densities to other
 quantities, so let us evaluate the
 time derivatives of them in terms of the moments.

Before that, note that for the time derivatives of any integral which includes the mass density

\be \dfrac{d}{dt} \int F(x,t) \rho (x,t) d^3x = \int \left[\dfrac{\partial F(x,t)}{\partial t} \rho(x,t) + F(x,t) \dfrac{\partial \rho(x,t)}{\partial t} \right]d^3x \label{ddt}. \ee
Using (\ref{mc}), and integrating by parts (keeping in mind that mass density vanishes on the boundary); the right-hand side of (\ref{ddt}) becomes

$$ \int \left[\dfrac{\partial F}{\partial t} \rho + (\dfrac{\partial F}{\partial x^a})\dot{x}_a \rho \right] d^3x $$

and since

$$ \dfrac{d}{dt} = \dfrac{\partial}{\partial t} + \dot{x^a}\dfrac{\partial}{\partial x^a} , $$

one has

\be \dfrac{d}{dt} \int F(x,t) \rho (x,t) d^3x = \int \dfrac{dF(x,t)}{dt} \rho(x,t) d^3x . \label{damourf}\ee
Using this result \cite{damour}, let us evaluate the time derivatives of mass and momentum density moments \cite{dixonIV,dixonV}.

For mass moments, let us consider lower moments separately:

For $n=0$:

\be m= \int\rho d^3x \rightarrow \dfrac{dm}{dt} =0,  \label{m1}\ee
which basically states the mass conservation.

$n=1$:

\be	m_a=\int r_a \rho d^3x \rightarrow \dfrac{dm_a}{dt} = \dfrac{d}{dt} \int r_a \rho d^3x = \int \rho \dfrac{dr_a}{dt} d^3x =  p_a - mv_a, 	 \label{pa}\ee
which is just the relative momentum with respect to the moving origin.

$n \geq 2$:

$$m_{a_1 a_2 ... a_n}= \int r_{a_1} r_{a_2} ... r_{a_n} \rho d^3x , $$

$$ \rightarrow \dfrac{d}{dt}m_{a_1 a_2 ... a_n}  = \int \dfrac{d}{dt}(r_{a_1} r_{a_2} ... r_{a_n}) \rho d^3x = n\int \dot{r}_{(a_1} r_{a_2} ... r_{a_n)} \rho d^3x, $$

\be \rightarrow \dfrac{1}{n} \dfrac{d}{dt}m_{a_1 a_2 ... a_n} = p_{({a_1} {a_2} ... {a_n})} - v_{(a_1} m_{{a_2} ... {a_n)}}. \label{pn}\ee
The physical meaning of these higher moments are not that transparent but we shall use them in our computations.
Let us find the analogous expressions for the momentum density moments.

$n=0$:

$$ p_a = \int \rho \dot{x_a} d^3x , $$

\be \dfrac{d}{dt}p_a =\dfrac{d}{dt}\int \rho \dot{x_a} d^3x = \int \dfrac{\partial} {\partial t}(\rho \dot{x_a}) d^3x =\int\left[-\dfrac{\partial}{\partial x^b}(\rho
  \dot{x}_a\dot{x}_b - \sigma_{ab} ) + \rho \dfrac{\partial}{\partial x^a}
  \phi\right]
  d^3x , \ee
  where we have made use of (\ref{cem}) and (\ref{damourf}).

Since the total divergence term vanishes on the boundary:

\be \dfrac{d}{dt}p_a =\int \rho \dfrac{\partial}{\partial x^a} \phi d^3x , 	 \label{p1}\ee
which is just the total gravitational force on the body.

$n=1$:

$$\dfrac{d}{dt} p_{ab} = \dfrac{d}{dt} \int \rho r_a \dot{x}_b d^3x. $$

In order to evaluate this, use (\ref{mc}) in (\ref{cem}) to get

$$\rho \dfrac{d}{dt} \dot{x}_a = \dfrac{\partial}{\partial x^b} \sigma_{ab} + \rho\dfrac{\partial}{\partial x^a} \phi. $$

Using this, one has:

$$\dfrac{d}{dt} p_{ab} = \int \rho \left [ (\dfrac{d}{dt}r_a) \dot{x}_b +  r_a \dfrac{d} {dt}\dot{x}_b \right]  d^3x = \int \left[\rho \dot{x}_a \ua_b -\rho v_a \ua_b +r_a(\dfrac{\partial}{\partial x^c}\sigma_{bc} + \rho \dfrac{\partial}{\partial x^b}\phi)\right]d^3x . $$

Since $r_a = x_a - z_a$, the term with the derivative of the stress tensor on the right hand side can be integrated by parts to give $-\int \delta_{ac} \sigma_{bc}d^3x .$

Then,

\be \dfrac{d}{dt} p_{ab} = \int \left[ -\rho v_a \ua_b + (\rho \dot{x}_a \ua_b - \sigma_{ba} ) + r_a  \rho \dfrac{\partial}{\partial x^b}\phi\right]  d^3x = -v_ap_b + t_{ab} + \int r_a \rho\dfrac{\partial}{\partial x^b}\phi d^3x . \label{p2}\ee
And finally, for $n \geq 2$, a similar computation gives

\be \dfrac{1}{n}\dfrac{d}{dt} p_{a_1 a_2 ... a_n b} = -v_{(a_1}p_{a_2 ... a_n)b} + t_{(a_1 ... a_n)b} + \dfrac{1}{n} \int \rho r_{a_1} ... r_{a_n} \dfrac{\partial} {\partial x^b}\phi d^3x . \label{p3}\ee
Note that the resultant infinite set of equations (\ref{m1}) - (\ref{p3}) are completely equivalent to the equations (\ref{mc}) and (\ref{cem}). All the information can be obtained from the moments, except the terms involving the gravitational potential. But these terms can be written in terms of mass moments by expanding $\dfrac{\partial}{\partial x^a}\phi$ about $z^a$:

$$\dfrac{\partial}{\partial x^a}\phi(x) = \sum_{n=0}^{\infty} \dfrac{1}{n!}r_{b_1}... r_{b_n} \dfrac{\partial}{\partial x^{b_1}}...\dfrac{\partial}{\partial x^{b_n}}\bigg( \dfrac{\partial}{\partial x^a}\phi(x)\bigg) \bigg|_{\vec{x}=\vec{z}}.$$

This shows that the moments $m_{a_1 a_2 ... a_n}, p_{a_1 a_2 ... a_n b}$ and $t_{a_1 a_2 ... a_n b c}$ , with the above infinite set of interrelations completely define the system.

Dixon's idea is to find a subset of these moments that defines the same system of equations, but subject to a finite number of interrelations. This can be done by finding the irreducible symmetries of the moments $p_{a_1 a_2 ... a_n b}$ and $t_{a_1 a_2 ... a_n b c}.$
\subsection{Reduced Moments}
It is simple to find the reduced moments of the momentum density. The defining equation (\ref{pm}), shows that these moments are symmetric in the first $n$ indices, and does not have a particular symmetry in the last index. This object can be separated into two irreducibly symmetric parts; one of which is totally symmetric in all indices, and the other antisymmetric in last index and each of the first $n$ indices. But it is better to find the irreducible symmetries in a more formal way that can also be used in the evaluation of irreducible symmetries of $t_{a_1 a_2 ... a_n b c}$'s.

  $p_{a_1 a_2 ... a_n b}$ is a tensorial object which is obtained by taking the
  tensor product of a rank $1$ tensor with a totally symmetric tensor of rank $n$.
  In order to evaluate the irreducible symmetries of it we will use representation theory of the symmetric group \cite{weyl}. Let $[\lambda_1,\lambda_2... \lambda_n]$ with $\lambda_1 \geq \lambda_2 \geq ...\geq \lambda_n$ correspond to the irreducible symmetry of the Young diagram with partition $(\lambda_1, \lambda_2 ..., \lambda_n)$ (see figure (\ref{fig:figu})).

  In order to evaluate the irreducible symmetries of $p_{a_1, a_2... a_n b}$, we need to consider the product of 2 Young diagrams with partition $(n)$ and partition $(1)$ \cite{dixonV}. Following the rules of multiplying the Young diagrams \cite{sakurai}, one has
  $[n]\otimes[1]=[n+1]\oplus[n,1]$.

   This shows that $p_{a_1 a_2 ... a_n b}$ can be written in terms of 2 irreducible parts as

   $$p'_{a_1 a_2 ... a_n b} = p_{(a_1 a_2 ... a_n b)}, \quad \quad \quad
   n\geq0,$$

   a totally symmetric part; and

   $$p''_{a_1 a_2 ... a_n b} = p_{(a_1 a_2 ... a_{n-1}[a_n) b]}, \quad \quad \quad
   n\geq0 . $$

   Instead of these $p''$'s, one can also use the following
   \vspace{3 em}
 \newcommand{\rma}{\mbox{$\lambda_1$}}
\newcommand{\rmb}{\mbox{$\lambda_2$}}
\newcommand{\rmc}{\mbox{$\lambda_n$}}
\vspace{3 em}
\begin{figure}[h!]
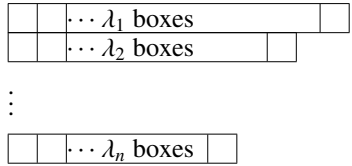

 \begin{center}
$\begin{array}{l}
\young(\,\,{\cdots \space \rma \space boxes \space \space \space \space \space \space \space \space \space \space \space \space \space \space \space \space \space \space \space }\hfil,\,\,{\cdots \space \rmb \space boxes \space \space \space \space \space \space \space \space \space \space \space }\hfil)\\
$\vdots$\\
\young(\,\,{\cdots \space \rmc \space boxes \space \space}\hfil)
\end{array}$\end{center}
\caption{Young diagram with partition $(\lambda_1, \lambda_2 ..., \lambda_n)$.}
\label{fig:figu}
\end{figure}
\vspace{3 em}
   $$p''_{a_1 a_2 ... a_n b c} = p_{a_1 a_2 ... a_{n}[b c]}, \quad \quad \quad
   n\geq0,$$

which satisfies

   $$p''_{a_1 a_2 ... a_n b c} = p''_{(a_1 a_2 ... a_{n})[b c]}, \quad \quad \quad
   n\geq0,$$

   $$p''_{a_1 a_2 ... [a_n b c]} = 0,  \quad \quad \quad n\geq1 . $$

One can see by checking (\ref{p3}) that $p'$'s are totally determined by $m$'s. The only moment left undetermined by this process is $p_a$, but it can be read directly from (\ref{pa}), so this moment is also determined by $m$'s. So $m$'s and $p''$'s can be used to determine the moments of the momentum density instead of $m$'s and $p$'s. The advantage of using the former is that it has only one restriction, that is (\ref{m1}), on the other hand the latter has to obey an infinite set of interrelations (\ref{m1}) - (\ref{pn}). $p''$'s are called ``Reduced moments" of $\rho \ua^a$ - the momentum density \cite{dixonV}.

The same can be done for the moments of the stress tensor. Using the rules for taking the direct product of Young Diagrams, it can be found that $t_{a_1 a_2 ... a_n b c}$ has symmetry $[n]\otimes[2]=[n+2]\oplus[n+1,1]\oplus[n,2]$. (This is valid only for $n \geq 2$, so moments with $n < 2$ must be treated separately.)

Equation (\ref{p3}) relates $t_{(a_1 a_2 ... a_n) b}$ to moments of mass and momentum density. Then, totally symmetric part of $t$'s (with symmetry $[n+2]$) and the part which is antisymmetric in 2 indices (the part with symmetry $[n+1,1]$) can be determined from $m$'s and $p$'s, or from $m$'s and $p''$'s.

The remaining part of $t$'s, with symmetry $[n,2]$ can be taken as

$$t'_{a_1 a_2 ... a_n b c} = t_{(a_1 a_2 ... a_{n-2} [a_{n-1} [a_n)( b] c)]}, \quad \quad \quad n\geq 2.$$

Equivalently, one can introduce

\be t'_{a_1 a_2 ... a_n b c d e } = t_{a_1 a_2 ... a_n [b [d c] e]}, \quad \quad \quad n\geq0 \label{tprime} .\ee
(with antisymmetrization over the pairs $(b,c)$ and $(d,e)$), and impose

\be t'_{a_1 a_2 ... a_n b c d e } = t'_{(a_1 a_2 ... a_n) [bc] [d e]},\quad \quad \quad n\geq0, \label{tprime1}\ee

\be t'_{a_1 a_2 ... a_n b [c d e] }, \quad \quad \quad n\geq0,\label{tprime2}\ee

\be t'_{a_1 a_2 ... [a_n b c] d e }, \quad \quad \quad n\geq1.\label{tprime3} \ee
In order to finalize the evaluation of reduced moments of stress tensor density, one has to consider the case with $n < 2$.

For $n=1$, $t_{abc} = t_{a(bc)}=t_{(abc)} + t_{(bc)a}$ as it can be seen by writing down the terms explicitly. From (\ref{p3}), it can be seen that both terms are determined by $m$'s and $p''$'s.

For $n=0$, $t_{ab}= t_{(ab)}$ is determined by the symmetric portion of equation (\ref{p2}). So the antisymmetric part of (\ref{p2}) is not included in this reduction. Also, as previously noted, (\ref{p1}) is also not included by this reduction.

In summary, we've seen that instead of the complete set of moments, $m$'s, $p$'s and $t$'s with an infinite set of constraints, (\ref{m1})-(\ref{p3}); one can use the reduced set of moments: $m$'s, $p''$'s and $t'$'s, which also represent the system defined by (\ref{mc}) and (\ref{cem}), with only 2 constraints (\ref{p1}) and the antisymmetric part of (\ref{p2}) \cite{dixonV}.
\subsection{Point Particle Approximation}
In this section, equations (\ref{p1}) and the antisymmetric part of (\ref{p2}) will be examined in detail \cite{dixonIV}.

Working on one of the bodies, it is convenient to define a mass center for the body. Following the definition of mass center in Newtonian mechanics, we see that mass center can be defined by setting mass density dipole moment to zero. Using equation (\ref{pa}),

\be{m_a(t) =  \int \rho r_a dV = 0} . \label{cm1}\ee
Following this, again by (\ref{pa}), we have the usual momentum-velocity relation

\be{\dfrac{dm}{dt}=0 \rightarrow p_a=mv_a}.\label{cm2}\ee
This is the same relation as if we were working with a point particle, so this velocity can be interpreted as the velocity of the body.

In the next step, total force and torque acting on the body will be defined. The right hand side of (\ref{p1}) can be defined to be the force

\be{\dfrac{dp_a}{dt} \equiv F_a=\int \rho \dfrac{\partial}{\partial x^a} \phi d^3x. 	 \label{force}}\ee
For the total torque, first the antisymmetric part of equation (\ref{p2}) will be defined as the total angular momentum of the body with respect to $z_a(t)$. This can be seen by using the center of mass relation

\bea
{S_{ab}} \equiv 2p_{[ab]} = 2\int \rho r_{[a} \dot{x}_{b]}d^3x &=& 2\left[\int \rho x_{[a}\dot{x}_{b]}d^3x - \int \rho z_{[a}\dot{x}_{b]}d^3x\right] \\
&=& 2\left[\int \rho x_{[a}\dot{x}_{b]}dV - \int \rho z_{[a}v_{b]}d^3x\right],
\label{snm}
\eea
by (\ref{cm2}). It is seen that these terms are basically the internal angular momentum with respect to the fixed origin and the angular momentum of the center of mass with respect to the moving origin $z_a(t)$. The difference of these is the total angular momentum of the body with respect to $z_a(t)$. Now, evaluating the time derivative of this term

$$\dfrac{dS_{ab}}{dt}=2\dfrac{d}{dt}\int\rho r_{[a}\dot{x}_{b]}d^3x= 2\left[ \int \rho \dot {r}_{[a} \dot{x}_{b]} + \int r_{[a}(\rho \dot{x}\dot{)}_{b]}\right]d^3x.$$

The first term on the left hand side can be written as $2\int v_{[a}\rho \dot{x}_{b]}d^3x$. For the second term, using (\ref{cem}), one has

$$\int r_{[a}(\rho \dot{x}\dot{)}_{b]}d^3x = \int r_a\left(\rho \dfrac{\partial}{\partial x^b}\phi -\dfrac{\partial}{\partial x^c}(\rho \dot{x}_b \dot{x}_c - \sigma_{bc}) \right)d^3x.$$

Integrating by parts and dropping a boundary term, and using once again (\ref{cem}); the result can be written as

\be{\dfrac{dS_{ab}}{dt} = 2 \left[ \int \rho v_{[a} \dot{x}_{b]}d^3x + \int \rho r_{[a}\dfrac{\partial}{\partial x^{b]}}\phi d^3x \right]}.\ee
Here the first term on the right hand side is purely kinematical, and vanishes when $z_a(t)$ is chosen as the center of mass; so the other term can be interpreted as the total torque acting on the body: $L_{ab} \equiv 2\int \rho r_{[a}\dfrac{\partial}{\partial x^{b]}}\phi d^3x$. Then we have:

\be{\dfrac{dS_{ab}}{dt}} - 2p_{[a}v_{b]}= L_{ab}. \label{torque} \ee
The next step is to separate the self and external fields. For the self field of the $i^{th}$ body \cite{damour, dixonIV}

\be\nabla^2 \phi^{(i)} = -4\pi G \rho^{(i)}\label{sf},\ee
with the boundary condition $\phi^{(i)}\rightarrow 0$ as $r\rightarrow\infty.$

Define the difference field as

$$ \hat{\phi}^{(i)} \equiv \phi - \phi^{(i)},$$

By the linearity of (\ref{poisson}) and (\ref{sf}) one can say that the difference field for a body is equivalent to the external field. Also, note that by definition (\ref{force}), (\ref{torque}), force and torque are also linear, and can be separated in the same way.

\begin{figure}[h]
\centering
\includegraphics[width=10cm,height=9cm]{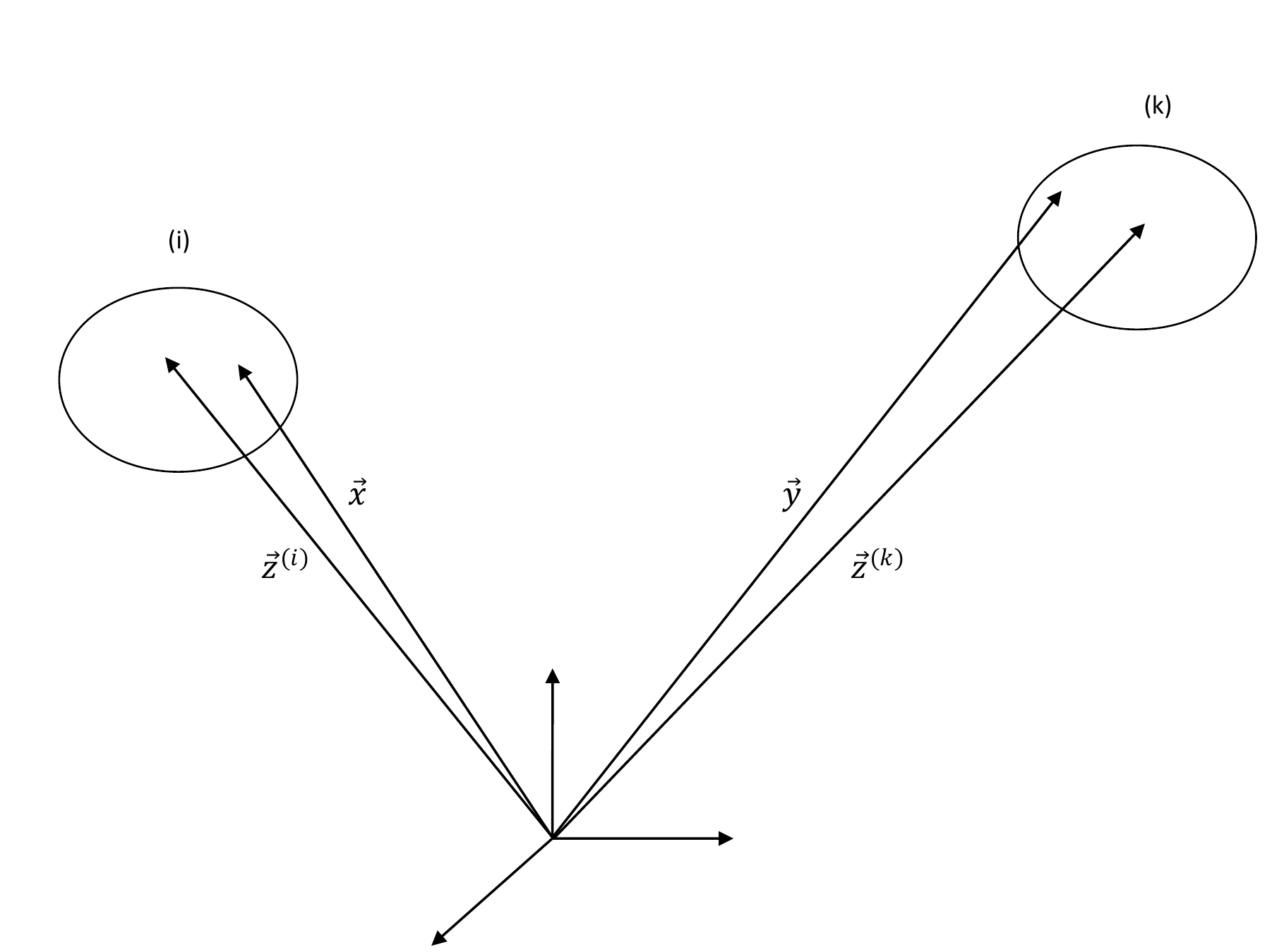}
\caption{Separation of the $i^{th}$ and the $k^{th}$ body.}
\label{fig:figure1}
\end{figure}

Self force and self torque can be evaluated explicitly as follows. The solution of Poisson's equation (\ref{poisson}) is

\be{\phi^{(i)}(\vec{x}, t) = G \int \rho^{(i)}(\vec{y}, t) \dfrac{1}{|\vec{x} - \vec{y}|} d^3\vec{y} ,   }\ee
and by (\ref{force}), the self force is given by

\be{F_a^{(i)} = \int \rho^{(i)}(\vec{x},t) \dfrac{\partial}{\partial x^a}\phi^{(i)}(\vec{x},t)d^3x    }= -G \int \int \rho^{(i)}(\vec{x},t) \rho^{(i)}(\vec{y},t)\dfrac{x_a - y_a}{|\vec{x} - \vec{y}|^3} d^3\vec{y} d^3\vec{x}, \ee
which vanishes because of the antisymmetry in $\vec{x}$ and $\vec{y}$ as expected by Newton's $3^{rd}$ law. Self torque also vanishes in the same way. So, all we have to consider is the external field acting on the particle. This can be written as

\be{\hat {\phi}^{(i)} = G \ \sum_{k\neq i} \int \rho^{(k)} (\vec{y},t) \dfrac{1}{|\vec{x} - \vec{y}|} d^3\vec{y}
       } \label{ext}.\ee
Here $\vec{y}$ denotes a point in the $k^{th}$ body
Now, let $z^{(k)}$ be a point in the $k^{th}$ body. Assuming that the separation between the bodies is large compared to the individual sizes of them (See figure (\ref{fig:figure1})), we have
\vspace{3 em}
\vspace{3 em}
\be |\vec{x} - \vec{z}^{(k)}| \gg |\vec{y} - \vec{z}^{(k)}|. \ee

Then we can expand $\dfrac{1}{|\vec{x} - \vec{y}|}$ in a Taylor series about $\vec{y} - \vec{z}^{(k)}$ as

\be \dfrac{1}{|\vec{x} - \vec{y}|} = \sum_{n=0}^\infty \dfrac{1}{n!}r_{a_1}^{(k)}...r_{a_n}^{(k)} \dfrac{\partial}{\partial y^{a_1}}...\dfrac{\partial}{\partial y^{a_n}} \dfrac{1}{|\vec{x} - \vec{y}|}    \bigg|_{\vec{y} = \vec{z}^k}, \ee
where $r_a^{(k)} = y_a - z_a^{(k)}.$

Then the external field can be expressed as

$$ \hat {\phi}^{(i)} = G  \sum_{k\neq i} \sum_{n=0}^\infty \int \dfrac{1}{n!}\rho^{(k)} (\vec{y},t)  r_{a_1}^{(k)}...r_{a_n}^{(k)} \bigg( \dfrac{\partial}{\partial y^{a_1}}...\dfrac{\partial}{\partial y^{a_n}} \dfrac{1}{|\vec{x} - \vec{y}|}\bigg) \bigg|_{\vec{y} = \vec{z}^{k}}    d^3{\vec{y}}.     $$

Using the definition of the mass density moments (\ref{mm}), this can be written in terms of the moments of the $k^{th}$ body

\be \hat {\phi}^{(i)} = G  \sum_{k\neq i} \sum_{n=0}^\infty  \dfrac{1}{n!} m^{(k)}_{a_1 ... a_n} \dfrac{\partial}{\partial y^{a_1}}...\dfrac{\partial}{\partial y^{a_n}} \dfrac{1}{|\vec{x} - \vec{y}|}    \bigg|_{\vec{y} = \vec{z}^{k}} .  \label{fii}  \ee
In order to evaluate the external force and the torque acting on the body, we can expand $\dfrac{\partial}{\partial x^a}\phi^{(i)}$ in a Taylor series about $z^{(i)}$ in the definitions (\ref{force}) and (\ref{torque}), once again using our starting assumptions. Then the force can be written as

\be \hat{F}_a^{(i)} = \int \rho \dfrac{\partial}{\partial x^a}\hat{\phi} dV = \int \rho \sum_{n=0}^{\infty}\dfrac{1}{n!} r_{b_1}... r_{b_n} \dfrac{\partial}{\partial x^{b_1}}...\dfrac{\partial}{\partial x^{b_n}} \dfrac{\partial}{\partial x^{a}} \hat{\phi}^{(i)} \bigg|_{\vec{x} = \vec{z}^(i)}  d^3x      , \ee
where $r_{b} = x_b - z^{i}_b$ as usual. Again using (\ref{mm})

\be \hat{F}_a^{(i)} =  \sum_{n=0}^{\infty} \dfrac{1}{n!} m^{(i)}_{ b_1 ... b_n} \dfrac{\partial}{\partial x^{b_1}}...\dfrac{\partial}{\partial x^{b_n}} \dfrac{\partial}{\partial x^{a}} \hat{\phi}^{(i)}\bigg|_{\vec{x} =\vec{z}^(i)}       , \ee
and similarly for $L_{ab} $

\be \hat{L}^{(i)}_{ab} = \sum_{n=0}^\infty \df{2}{n!} m^{(i)}_{[a|b_1...b_n} \df{\partial}{\partial x^{b_1}}...\df{\partial}{\partial x^{b_n}} \df{\partial}{\partial x^{|b]}} \hat{\phi}^{(i)}\bigg|_{\vec{x} = \vec{z}^{(i)}}.\ee
In transition to the point particle approximation, only the first terms in the series of $\hat{\phi}^{(i)}$ and $\hat {F}^{(i)}_a$ will be taken. (The accuracy of the approximation could be maximized by taking $z^{(k)}$ to be center of mass of the $k^{th}$ body, so that the second term in each series identically vanishes.)

Expanding (\ref{force})

$$\dfrac{dp_a}{dt} = F_a,$$
in the center of mass frame of the $i^{th}$ body (the body we are working on), we have the momentum velocity relation (\ref{cm2}). Then

$$m^{(i)} \dfrac{d^2z_a}{dt^2} = \hat{F}^{(i)}_a= m^{(i)}\dfrac{\partial}{\partial z^a}\hat{\phi}^{(i)}= m^{(i)} G\sum_{k \neq i} m^{(k)} \bigg( \dfrac{\partial}{\partial x^a} \bigg (\dfrac{1}{|\vec{x} - \vec{z}^{(k)}|} \bigg) \bigg|_{\vec{x} = \vec{z^{(i)}}} ,      $$

\be \dfrac{d^2z_a}{dt^2} = -G\sum_{k \neq i} m^{k} \dfrac{(z^{(i)} - z^{(k)})_a}{|\vec{z}^{(i)} - \vec{z}^{(k)}|^3},\ee
which is just the inverse square force law. This equation can be solved for given initial conditions and known masses. It is independent of the internal structure of the bodies \cite{damour, dixonIV}.
\subsection{Internal Structure and Higher Moments}
Going one step further, and including the second order terms in each series, internal structure of the bodies can be accounted for in the equation of motion. Using the center of mass condition

\be
m^{(i)}\dfrac{d^2 z^{(i)}_a}{dt^2} = \hat{F}^{(i)}_a = m^{(i)}
\bigg( \df{\partial}{\partial x^a} \hat{\phi}^{(i)}\bigg)_{\vec{x}  \vec{z}^{(i)}}
+  m^{(i)}_{b_1 b_2} \bigg( \df{\partial}{\partial x^{b_1}}\df{\partial}{\partial
x^{b_2}}\df{\partial}{\partial x^a}\hat{\phi}^{(i)}\bigg)_{ \vec{x} =
 \vec{z}^{(i)}} + \cdots \label{2ndorder}\quad \ee
$2^{nd}$ order mass density moment is the inertia tensor

\be \mi_{b_1 b_2} = \int \rho r_{b_1}r_{b_2} d^3x. \ee
By (\ref{ext}) and (\ref{poisson}), the Laplacian of the external field is zero within the body that is being dealt with, one can equivalently use the traceless `Quadrupole Tensor' instead of the inertia tensor \cite{damour}

\be Q^{(i)}_{b_1 b_2} = \mi_{b_1 b_2} - \dfrac{1}{3} \delta_{b_1 b_2} \mi_{b_1 b_2}. \label{newquad}\ee
This does not change anything in (\ref{2ndorder}), but the fact that it can be written in this form shows that the point particle approximation, for bodies whose trace of the inertia tensor has a large norm, has greater accuracy.

Now, using the expansion of $\fii$ as in (\ref{fii})

\be \fii(\vec{x},t) = G \sum_{k\neq i} \bigg[ m^{(k)} \bigg( \df{1}{|\vec{x} - \vec{y}|}\bigg)_{\vec{y} = \vec{z}^{(k)}} + m^{(k)}_{a_1 a_2} \bigg( \df{\partial}{\partial y^{a_1}}\df{\partial}{\partial y^{a_2}} \df{1}{|\vec{x} - \vec{y}|}\bigg)_{\vec{y}=\vec{z}^{(k)}}+...\bigg]      ,\ee
again introducing the quadrupole tensor, force can be written as

\be
\begin{split}
m^{(i)}\dfrac{d^2 z^{(i)}_a}{dt^2} &= G \sum_{k \neq i} \bigg{\{}    \mi \mk \bigg[ \df{\partial}{\partial x^a} \bigg( \df{1}{|\vec{x} - \vec{y}|}\bigg)_{\vec{y} = \vec{z}^{(k)}}\bigg]_{\vec{x} = \vec{z}^{(i)}}
  +
 \df{1}{2} \mi Q^{(k)}_{a_1 a_2} \bigg[ \df{\partial}{\partial x^a} \bigg( \df{\partial}{\partial y^{a_1}}\df{\partial}{\partial y^{a_2}} \df{1}{|\vec{x} - \vec{y}|}\bigg)_{\vec{y} = \vec{z}^{(k)}}\bigg]_{\vec{x} = \vec{z}^{(i)}}
 \\ &+
 \df{1}{2}\mk Q^{(i)}_{b_1 b_2} \bigg[  \df{\partial}{\partial x^{b_1}}\df{\partial}{\partial x^{b_2}}\df{\partial}{\partial x^a} \bigg( \df{1}{|\vec{x} - \vec{y}|}\bigg)_{\vec{y} = \vec{z}^{(k)}}\bigg]_{\vec{x} = \vec{z}^{(i)}}  +  ... \bigg{\}}
  .
  \end{split} \ee
  This can be written in a more compact form

   \be m^{(i)}\dfrac{d^2 z^{(i)}_a}{dt^2} = G \sum_{k \neq i} \bigg[ \mi \mk \df{\partial}{\partial z^{(i)a}} \bigg( \df{1}{|\vec{z}^{i} - \vec{z}^{k}|}\bigg) + \df{1}{2} (m^{(i)}Q^{(k)}_{a_1 a_2} + m^{(k)}Q^{(i)}_{a_1 a_2} )\df{\partial^3}{\partial z^{(i)a} \partial z^{(i)a_1} \partial z^{(i)a_2}}\bigg( \df{1}{|\vec{z}^{i} - \vec{z}^{k}|}\bigg) +...\bigg],\ee
   where we have used

   \be \df{\partial^2}{\partial y^{a_1} \partial y^{a_2}} \df{1}{|\vec{x} - \vec{y}|} = \df{\partial^2}{\partial x^{a_1} \partial x^{a_2}} \df{1}{|\vec{x} - \vec{y}|}.\ee
If we let $L$ be the typical length of one of the bodies and $R$ the typical separation between each body; the first term in the above series is of the order $R^{-2}$, the second and the third is $(L^2/R^4)$. The first neglected term (dipole-dipole interaction) is of the order $(L^4/R^6)$.\\ In contrast to the point particle case, as higher moments is considered, the motion becomes indeterminate since there is no information on the time dependence of these moments. To get rid of this, now it will be assumed that the bodies are rigid objects \cite{dixonIV}.
\subsubsection{Rigid Objects}
In Newtonian mechanics, an object is called rigid if the relative separation of the particles forming the body does not change over time \cite{landaumech}. Consider the infinitesimal motion of a point in such a body. Such a motion, $d\vec{x}$, with respect to fixed origin can be written in two parts. One is a part that is equal to the movement of the center of mass $d\vec{z}$, and the other is the motion of the corresponding particle with respect to center of mass. We are assuming that the separation of these two points don't change, so it can only include a rotation of the point $r_a$ around the center of mass: $d\vec{\phi} \wedge \vec{r}$, where $d\phi$ is the angle of rotation. Then,

\be d\vec{x} = d\vec{z} + d\vec{\phi} \wedge \vec{r},\ee
from which, after dividing by $dt$, one obtains

\be \vec{\dot{x}} = \vec{\dot{z}} + \vec{\Omega} \wedge \vec{r}, \ee
where $\vec{\Omega} \equiv \df{d\vec{\phi}}{dt}$, and we have used the assumption that the relative position of the particles to be constant \cite{landaumech}. This can be expressed as

\be{\ua_a(\vec{x}, t) = \dot{z}_a(t) - \Omega_{ab}(t) r_{b}(\vec{x},t)        }\ee
using index notation, for some antisymmetric $\Omega_{ab}(t)$.

Using this in (\ref{pm}), one obtains \cite{dixonIV}

$$ p_{a_1 ... a_n}b = m_{a_1 ... a_n} \dot{z}_b + \Omega_{cb}m_{a_1 ... a_n c}. $$

 Since $z_a(t)$ is the center of mass, using the vanishing of the mass dipole moment, we have

\be{S_{ab} = p_{[ab]} = \Omega_{c[b}m_{a]c}}. \label{sr}\ee
And the time dependence of any $m$ can be found as

$${\dfrac{dm_{a_1 ... a_n}}{dt} = \int \dfrac{dr_{a_1}...r_{a_n}}{dt} \rho d^3x = \int n (\ua-v)_{(a_1}r_{a_2}... r_{a_n)} \rho d^3x = \int n r_c \Omega_{c(a_1} r_{a_2} ... r_{a_n)} \rho d^3x            },$$

\be{\dfrac{dm_{a_1 ... a_n}}{dt} = n \Omega_{c(a_1} m_{a_2 ... a_n)c}}.\ee
With these in hand, (\ref{cm2}) and (\ref{sr}) can be used to eliminate momentum and angular momentum from (\ref{force}) and (\ref{torque}). Resulting equations are

\be m\df{dv_a}{dt} = \int \rho \df{\partial}{\partial x^a} \phi d^3x, \ee

\be \bigg( \dot{\Omega}_{c [ b} + \Omega_{cd}\Omega_{d [b}\bigg)m_{a]c} = \int \rho r_{[a} \df{\partial}{\partial x^{b]}} \phi d^3x.\ee
Together with the multipole expansion of the force and the torque, we have a determinate set of equations for $v_a$, $\Omega_{ab}$ and $m_{a_1 ... a_n}$, irrespective of the order of moments kept in the calculation \cite{dixonIV}.

\section{GENERAL RELATIVITY}
\label{sec:General Relativity}

In order to discuss the problem of motion of a system of bodies in General Relativity, relativistic versions of  (\ref{mc}), (\ref{cem}), and (\ref{poisson}) are needed. In this case, equations of motion are \cite{dixonIV} (replacing the continuum equation of motion in Newtonian Theory)

\be R_{\alpha \beta} - \dfrac{1}{2}Rg_{\alpha \beta} = \kappa T_{\alpha \beta}, \label{ee}\ee

and equations (\ref{mc}), (\ref{cem}) will be replaced by

\be \nabla_\beta T^{\alpha \beta} = 0, \label{eom}\ee

which is obtained by the contracted Bianchi identity. These set of equations fully describe the motion, and the aim is to follow the procedure followed in the previous chapter to extract them in a convenient form. The difficulty comes from the nonlinearity of the equations. In the Newtonian problem (due to linearity of the theory) it was possible to separate the self field and the external field for one of the bodies, but this is no longer the case in a relativistic theory. Also, in Newtonian theory, gravitational force is totally determined only by mass density, while in General Relativity all forms of energy have some gravitational mass, so if we were to make a similar moment expansion as in (\ref{mm}), this will not be an expansion only for $T^{00}$, but also for the other components of the energy momentum tensor \cite{dixonIV, dixonIII}.

In the previous chapter, the first step was to describe a mass center for the body in concern. Here, however, first the definitions of momentum and angular momentum will be given, since the center of mass definition that fits in with the Dixon's theory involves these definitions.

\subsection{Momentum and Angular Momentum}

The aim of this section is to define momentum and angular momentum of a spatially bounded body, which is described by some symmetric energy momentum tensor $\emt$. For a finite sized object, $\emt$ is nonzero in the world tube $W$ of the body, which extends to past and future infinity but is spatially bounded. Assume, for now, that the spacetime admits isometries described by a Killing vector field $\xi$, such that

\be\nabla_{(\alpha} \xi_{\beta)} =0. \label{kve}\ee

In such a spacetime, there exists some constants of motion resulting from the symmetries  of the spacetime, which can be obtained by contracting the energy momentum tensor with the Killing vector field \cite{de felice}. By (\ref{kve}) and (\ref{eom}), and the symmetry of  $\emt$

\be \nabla_\beta(\xi_\alpha T^{\alpha \beta}) = (\nabla_\beta \xi_\alpha) T^{\alpha \beta}+ \xi_{\alpha}(\nabla_\beta \emt) = 0       . \ee

\begin{figure}[h]
\centering
\includegraphics[height=8cm]{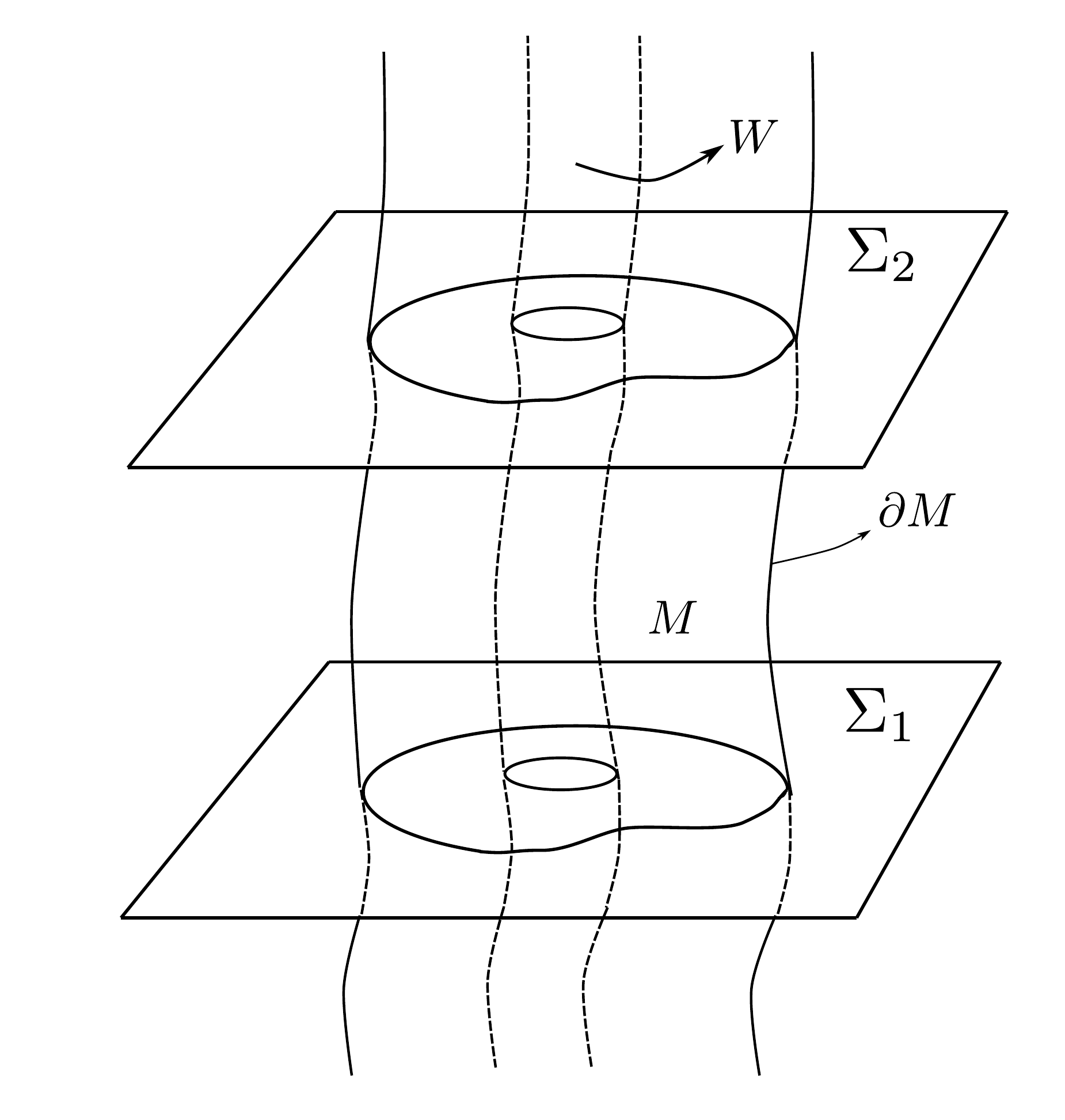}
\caption{Region of integration.}
\label{fig:figure2}
\end{figure}

Integrate this over some volume $M$ which includes some portion of the object`s world-tube $W$; whose surface is everywhere timelike and bounded by two space-like, non-intersecting hypersurfaces $\Sigma_1$ and $\Sigma_2$, such that $\Sigma_2$ is the future of $\Sigma_1$. Choose $M$ such that $\emt$ vanishes on the surface of it (see figure (\ref{fig:figure2})).

\be  \int_{M} \sqrt{-g} \nabla_{\alpha}(\xi_{\beta}\emt) d^4x =0  ,         \ee

which can also be written as

\be \int_{M} \partial_{\alpha} (\sqrt{-g} \xi_{\beta} \emt) d^4x =0. \ee

Using Stoke's theorem, this can be written as an integral over the boundary of the volume

\be  \int_{\partial M}   \xi_{\beta} \emt d\Sigma_\alpha =0   .  \ee

 Writing the contributions of each boundary of $M$ separately
 \vspace{3 em}
\vspace{3 em}
\be \int_{\Sigma_2 \cap M}   \xi_{\beta} \emt d\Sigma_\alpha + \int_{\Sigma_1 \cap M}  \xi_{\beta} \emt d\Sigma_\alpha + \int_{\partial M}   \xi_{\beta} \emt d\Sigma_\alpha = 0. \ee

The last term vanishes since $\emt$ is zero on $\partial M$, and the first two terms can be restricted to $\Sigma_{\alpha} \cap W$ by the same reasoning. Then,

 $$ \int_{\Sigma_2 \cap W}   \xi_{\beta} \emt d\Sigma_\alpha = - \int_{\Sigma_1 \cap W}   \xi_{\beta} \emt d\Sigma_\alpha.  $$

 So,

 \be  \int_{\Sigma}   \xi_\alpha \emt d\Sigma_\beta = C\ee

 is a constant of motion, independent of the hypersurface chosen \cite{dixonIV, de felice}.

After this construction, we will define the momentum and angular momentum of this system in such a way that the symmetries of the spacetime will lead to a vanishing linear combination of them, and they will match with the known definitions in special relativistic limit. Using (\ref{egds}),

 \be  \int_{\Sigma}   ({K_\alpha}^{\kappa} \xi_{\kappa} + {H_\alpha}^{\kappa} \sigma^{\lambda} \nabla_{[\kappa} \xi_{\lambda]})\emt d\Sigma_\beta = C . \ee

   Here $\xi_\kappa = \xi_\kappa(z)$ and $\nabla_{[\kappa} \xi_{\lambda]}= \nabla_{[\kappa} \xi_{\lambda]}(z)$, and $z$ is a fixed but otherwise arbitrary point.
 Define momentum and angular momentum as \cite{dixonIV, dixonI}

 \be p^\kappa(z,\Sigma) \equiv \int_{\Sigma} \emt {K_\alpha}^\kappa d\Sigma_\beta , \label{mom}          \ee

 \be  S^{\kappa \lambda}(z, \Sigma) \equiv 2\int_\Sigma \emt {H_\alpha}^{[\kappa}\sigma^{\lambda]}d\Sigma_\beta . \label{angmom}\ee

The constant can be written as

\be  C = p^{\kappa}(z,\Sigma) \xi_\kappa + \dfrac{1}{2}  S^{\kappa \lambda}(z, \Sigma) \nabla_{[\kappa} \xi_{\lambda]} . \label{cons}    \ee

Since the definitions of $p^\kappa$ and $S^{\kappa \lambda}$ do not depend on the  Killing vector fields, the definitions can be used for an arbitrary spacetime without any symmetries. However, when the spacetime admits isometries, $C$ gives a linear combination of momentum and angular momentum which is a constant \cite{dixonIV, dixonI}.

In the flat space limit, $\sigma^\kappa$ is equivalent to the displacement vector of $-(x-z)^\kappa$ as is shown in (\ref{skf}), and ${K^\alpha}_\lambda$ and ${H^\alpha}_\lambda$ both reduce to unit tensors \cite{poisson}, so the special relativistic limit of the above momentum and angular momentum are

\be p^\kappa = \int T^{\kappa \beta} d\Sigma_\beta, \label{psr}\ee
\be S^{\lambda \kappa} = 2 \int (x-z)^{[\lambda} T^{\kappa]\beta} d\Sigma_\beta\label{ssr}, \ee

which are the usual definitions obtained by using Noether's theorem in classical field theories \cite{dixonV}. This result supports the validity of the given definitions, (\ref{mom}) and (\ref{angmom}), in an arbitrary spacetime.

\subsection{Force and Torque}
$p^\kappa$ and $S^{\kappa \lambda}$ as given above depend on the fixed but otherwise arbitrary point $z$ and the hypersurface of integration $\Sigma$, but (\ref{cons}) is independent of both. So one can vary $z$ along a parametrized worldline $z(s)$, and use the constancy of (\ref{cons}) along it to get
$$\df{\delta C}{\delta s} = 0, $$
where $s$ is the parameter along $z$. Then we have
\be  \df{\delta}{\delta s}\bigg[ p^\kappa \xi_\kappa + \df{1}{2}S^{\kappa \lambda} \nabla_{[\kappa} \xi_{\lambda ]}\bigg] = 0. \ee
Or writing the terms explicitly
\be \df{\delta p^\kappa}{\delta s} \xi_\kappa + p^\kappa \df{\delta}{\delta s} \xi_\kappa +\df{1}{2}\bigg( \df{\delta S^{\kappa \lambda}}{\delta s}  \nabla_{[\kappa} \xi_{\lambda ]} + S^{\kappa \lambda} \df{\delta}{\delta s} \nabla_{[\kappa} \xi_{\lambda ]}     \bigg) = 0. \ee

By (\ref{kve}), we have $\df{\delta}{\delta s} \xi_\kappa = v^\lambda \nabla_{[\lambda} \xi_{\kappa ]}$ with $v^\lambda \equiv \df{d z^\lambda}{ds}$;
 and since a Killing vector field satisfies
\be  \nabla_\alpha \nabla_\beta \xi_\gamma=  R_{\beta \gamma \alpha \delta} \xi^\delta \label{kvk},\ee
the above result can be written as,
\be \df{\delta C}{\delta s} = \xi_\kappa \bigg[ \df{\delta p^\kappa}{\delta s} +
\df{1}{2}S^{\delta \lambda} v^\mu {R_{\delta \lambda \mu}}^\kappa   \bigg]
+\df{1}{2} \nabla_{[\kappa} \xi_{\lambda ]} \bigg[ \df{\delta S^{\kappa
\lambda}}{\delta s} - 2p^{[\kappa}v^{\lambda ]} \bigg] = 0 \label{con5}.\ee
This holds for all Killing vector fields if each term in the brackets separately vanishes. They will be defined to be the total force and total torque acting on the body \cite{dixonIV, dixonI}.

\be F^\kappa \equiv \df{\delta p^\kappa}{\delta s} + \df{1}{2}S^{\delta \lambda} v^\mu {R_{\delta \lambda \mu}}^\kappa   , \label{forcegr}\ee
\be L^{\kappa \lambda} \equiv  \df{\delta S^{\kappa \lambda}}{\delta s} - 2p^{[\kappa}v^{\lambda ]}, \label{torkcm}.\ee
With these definitions, (\ref{con5}) can be written as
\be \xi_\kappa F^\kappa + \df{1}{2}\nabla_{[\kappa} \xi_{\lambda ]} L^{\kappa \lambda}\label{cons1} = 0.\ee
Both definitions, (\ref{forcegr}) and (\ref{torkcm}), can be generalized to arbitrary spacetimes, since they don't depend on the Killing vectors. But in a general spacetime, higher multipole moments will also contribute to force and torque.  (\ref{cons1}) expresses the connection between the integrals of motion and the isometries of the spacetime. By (\ref{jacgensol}), it can be seen that the maximum number of linearly independent Killing vectors $\xi^\alpha (x)$ is equal to the maximum number of independent values of the initial values $\xi_\kappa(z),\nabla_{[\kappa}\xi_{\lambda]}(z)$ \cite{de felice}. So in a maximally symmetric spacetime, $F^{\kappa}$ and $S^{\kappa \lambda}$ vanish separately.

As a last note, keep in mind that by construction (\ref{cons1}) holds independent of the hypersurface chosen, but the force and torque as define above, depend on it. Also, in each definition, there is a term involving $v^\kappa$, the kinematical velocity, which will be defined in the next section to be the velocity of the center of mass \cite{dixonIV}.

\subsection{Center of Mass Worldline}

Before defining the multipole moments which carry all the information about the body, the definition of the center of mass worldline will be made. This will reduce the arbitrariness in the definitions of the worldline $z^\alpha$ and the hypersurface of integration $\Sigma$. This is the worldline with respect to which the multipole expansion will be made \cite{dixonIV, dixonI}.

Before giving the definition of center of mass in a curved spacetime, it will be
convenient to make the definition in the flat space of Special Relativity. In Newtonian mechanics, the definition of center of mass is done through the vanishing of the mass dipole moment, and relates the center of mass velocity to the momentum of the object by (\ref{cm2}). Also, note that by (\ref{snm}), only internal spin contributes in the center of mass frame. One can
directly generalize the definition made for Newtonian mechanics (\ref{cm1}) as
\cite{dixonI, de felice}

\be z^a \int_{x^0=constant } T^{00}d^3\Sigma_0 =  \int_{x^0=constant } T^{00} x^a d^3\Sigma_0      ,  \ee
where $a$ runs from 1 to 3. This can be put in a better form by introducing the notation $n_\alpha = \delta^0_\alpha$ as the unit normal to the hypersurface chosen, defining the Lorentz frame in which the body is at rest, and using $d^3\Sigma_\alpha= n_\alpha d^3\Sigma$ where $d^3 \Sigma$ is the invariant volume element \cite{de felice}
\be z^a \int_{x^0=constant } n_\alpha n_\beta T^{\alpha \beta} d\Sigma =\int_{x^0=constant } x^a n_\alpha n_\beta T^{\alpha \beta} d\Sigma,\ee
which can be written as
\be n_\alpha \int_{x_0 = constant} (z^a - x^a) T^{\alpha\beta} n_\beta d\Sigma = 0.\ee
Using (\ref{skf}), (\ref{psr}) and (\ref{ssr}), above condition can be written as
\be n_\alpha S^{\alpha \beta} = 0 \label{cmsr}.\ee
The left hand side of (\ref{cmsr}) will be our generalization of the mass dipole moment in a relativistic theory. It can be shown to reduce to the Newtonian definition (\ref{cm1}) in the limit. (\ref{cmsr}) can be carried to curved spacetime. In a general spacetime, we would like to have a momentum of the form
\be p^\kappa = M n^\kappa. \label{momvelgr}\ee
for a timelike vector $n^\kappa$ at $z$ with $M$ not necessarily a constant. But it would be natural to expect $M$ to reduce to a constant if the spacetime admits some symmetries. It has been shown in \cite{bb, schattnercm} that if the gravitational field varies slowly over the body, there exists a timelike future directed vector $n^\kappa$ such that
\be p^{[\lambda} n^{\kappa]} = 0.\ee
Generally, $n^\kappa$ will be different from $v^\kappa$ \cite{dixonIV, abrahamtez}, the tangent to the worldline and its normalization will be chosen as
\be n_\kappa v^\kappa = 1. \label{cmnorm}\ee
Hypersurface of integration can be constructed such that it is generated by all geodesics through $z$ orthogonal to $n^{\kappa}$. Using (\ref{wfz}), this tells that $\Sigma$ is the set of all $x$ such that
\be n_\kappa(z) \sigma^\kappa(z,x) = 0.\label{nsig}\ee
Then, the special relativistic definition of the mass center can be generalized to an arbitrary spacetime, using the definition of momentum as being parallel to $n^\kappa$, we have
\be p_{\kappa} S^{\kappa \lambda} = 0 \label{cmgr}.\ee
Uniqueness of such a worldline has been proven in \cite{bb, schattnercm}.
This construction removes the arbitrariness of the hypersurfaces of integration and of the central line $z^\kappa (s)$. Note that by (\ref{cmgr}), $S^{\kappa \lambda}$ has 3 linearly independent components \cite{dixonI}. This can be written explicitly by defining the spin vector as
\be S^{\kappa} = \df{1}{2} \eta_{\kappa \lambda \mu \nu} n^\nu S^{\lambda \mu}\label{spin},\ee
where $\eta_{\alpha \beta \gamma \delta} \equiv \sqrt{-g} \epsilon_{\alpha \beta \gamma \delta}$ is the volume $4$-form and $\epsilon_{\alpha \beta \gamma \delta}$ with $\epsilon_{0123} =1$ is the Levi-Civita symbol. (\ref{spin}) satisfies
\be n_{\kappa} S^\kappa =0. \ee

 $S^{\kappa \lambda}$ can be written in terms of this spin vector with the use of (\ref{cmgr}) as
\be S^{\kappa \lambda} = \eta^{\kappa \lambda \mu \nu} n_{\mu} S_\nu ,   \ee
which can be interpreted as the generalization of vanishing of the orbital component of the angular momentum in the center of mass frame \cite{abrahamtez}.

It is useful to study how $M$ in (\ref{momvelgr}) changes along $z^{\kappa}$. First of all, since $n^\kappa$ was defined to be a unit vector, it satisfies:
\be n_\kappa n^\kappa = 1 \quad, \quad \quad \quad n^\kappa \df{\delta n_\kappa}{\delta s} = 0 \label{nunit}.\ee
Using the definition (\ref{momvelgr}) in (\ref{forcegr}), one obtains
\be F^\kappa = \df{\delta M}{\delta s} n^\kappa + M \df{\delta n^\kappa}{\delta s} + \df{1}{2}S^{\delta \lambda} v^\mu {R_{\delta \lambda \mu}}^\kappa      \label{mdegisim}.  \ee
In flat space, $F^\kappa$ vanishes by \ref{cons1} and  $R_{\lambda \mu \nu}^\kappa = 0$ The contraction of the above equation with $n^\kappa$ shows, by (\ref{nunit}), that
\be \df{\delta M}{\delta s} = 0.\ee
Let us continue the problem in a general spacetime. Contracting (\ref{mdegisim}) with $v^\kappa$ gives
\be F^\kappa v_\kappa = \df{\delta M}{\delta s} n^\kappa v_\kappa  + M \df{\delta n^\kappa}{\delta s} v_\kappa + \df{1}{2}S^{\delta \lambda} v^\mu v_\kappa {R_{\delta \lambda \mu}}^\kappa.  \label{mdeggenel}\ee
The last term vanishes due to the antisymmetry of the Riemann tensor. By using (\ref{cmnorm}) and differentiating the center of mass condition in the form (\ref{cmsr}) along $z$
\be n_\kappa \df{\delta S^{\kappa \lambda}}{\delta s} = - \df{\delta n_\kappa}{\delta s} S^{\kappa \lambda}    .\ee
Contracting this with $\df{\delta n_\lambda}{\delta s}$ gives, by the antisymmetry of $S^{\kappa \lambda}$,
\be n_\kappa \df{\delta n_\lambda}{\delta s} \df{\delta S^{\kappa \lambda}}{\delta s} = 0.\ee
Contracting (\ref{torkcm}) with $n_\kappa \df{\delta n_\lambda}{\delta s}$ and using (\ref{cmnorm}) gives
\be L^{\kappa \lambda} n_{\kappa} \df{\delta n_\lambda}{\delta s} = -M v^\lambda \df{\delta n_\lambda}{\delta s}.\ee
Using this in (\ref{mdeggenel}), one has
\be \df{\delta M}{\delta s} = F^\kappa v_\kappa + L^{\kappa \lambda}n_\kappa \df{\delta n_\lambda}{\delta s}.\ee
So when force and torque vanish (which is the case in a maximally symmetric spacetime), $M$ is a constant \cite{de felice}.

An explicit solution of the center of mass velocity can also be found in terms of other variables as in \cite{ehlers}. In order to do this, one has to solve (\ref{forcegr}) and (\ref{torkcm}) in terms of $v^\kappa$ with constraints (\ref{momvelgr}), (\ref{cmnorm}) and (\ref{cmgr}) \cite{dixonIV}.
Before proceeding to the solution of the general problem, let us again evaluate the situation in flat spacetime, and work without the normalization $n_\kappa v^\kappa$ for now. In flat space, (\ref{forcegr}) and (\ref{torkcm}) reduce to
\be \df{\delta p^\kappa}{\delta s} = 0 \label{momconsr},\ee
and
\be \df{\delta S^{\kappa \lambda}}{\delta s} = 2p^{[\kappa}v^{\lambda ]}    \label{sdegsr}.\ee
Contract (\ref{sdegsr}) with $n_\kappa$ to get
\be n_\kappa \df{S^{\kappa \lambda}}{\delta s} =p^{\kappa } v^{\lambda} n_\kappa - p^{\lambda} v^{\kappa} n_\kappa.\ee
Using (\ref{momvelgr}), this can be written as
\be n_\kappa \df{S^{\kappa \lambda}}{\delta s} = M v^{\kappa}(\delta^\lambda_\kappa - n_\kappa n^\lambda)\label{sdegsr1}.\ee
The term in the parenthesis projects orthogonal to $n^\kappa$, and its contraction with $v^\kappa$ measures the velocity of the center of mass frame from the rest frame of an observer which has velocity $n_\kappa$ \cite{de felice}.
Now, using the constancy of momentum in flat space (\ref{momconsr}), differentiate the center of mass condition in the form (\ref{cmsr}) along with respect to $s$ to get
\be n_\kappa \df{\delta S^{\kappa \lambda}}{\delta s} = 0.\ee
This shows that the left hand side of (\ref{sdegsr1}) vanishes. Then one has
\be v^\lambda = n_\kappa v^\kappa n^\lambda\ee
That tells that $v^\kappa$ is parallel to $n^\kappa$, and identical to it when it is normalized as in (\ref{cmnorm}). Since $M$ was shown to be constant before, it is seen that $v^\kappa$ and $n^\kappa$ are also constant along $z$ in flat space. (\ref{cmsr}) with (\ref{momvelgr}) and the constancy of $M$ also show that the spin vector (\ref{spin}) is constant along $z$. This discussion shows that the center of mass worldline in flat space is a geodesic, and the spin vector is parallelly propagated along it \cite{de felice}.

Now its time to proceed to the aforementioned problem of finding the explicit solution of the center of velocity in terms of other parameters \cite{dixonIV, ehlers}. As a first step, differentiate (\ref{cmgr}) along $z$ to get
\be \df{\delta }{\delta s} (p_\lambda S^{\kappa \lambda}) = 0\rightarrow \df{\delta p_\lambda}{\delta s}S^{\kappa \lambda} + p_\lambda \df{\delta S^{\kappa \lambda}}{\delta s} = 0.\ee
Use (\ref{torkcm}) and (\ref{momvelgr}) in the second term to get
\be \df{\delta p_\lambda}{\delta s}S^{\kappa \lambda} + M n_\lambda \bigg[  L^{\kappa \lambda} + 2M n^{[\kappa} v^{\lambda ]} \bigg] = 0.\ee
Use (\ref{cmnorm})
\be \df{\delta p_\lambda}{\delta s}S^{\kappa \lambda} + Mn_\lambda L^{\kappa \lambda} + M^2 n^\kappa - M^2 v^\kappa = 0.\ee
For simplicity, let $Mn_\lambda L^{\kappa \lambda}+ M^2 n^\kappa \equiv M^2 t^\kappa$ so that the equation can be written in the form
\be \df{\delta p_\lambda}{\delta s}S^{\kappa \lambda} + M^2 [t^\kappa - v^\kappa] = 0 \label{t}.\ee
This can be inverted to get
\be v^\kappa = t^\kappa + \df{1}{M^2} \df{\delta p_\lambda}{\delta s} S^{\kappa \lambda}.\ee
Use this in (\ref{forcegr}) to get
\be \df{\delta p_\lambda}{\delta s} = F_\lambda - \df{1}{2M^2}S^{\sigma \mu} S^{\nu \tau} R_{\sigma \mu \nu \lambda} \df{\delta p_\tau}{\delta s} - \df{1}{2}S^{\sigma \mu} R_{\sigma \mu \nu \lambda} t^\nu.\ee
To get an equation for $\df{\delta p_\lambda}{\delta s} S^{\kappa \lambda}$, contract this with $S^{\kappa \lambda}$
\be S^{\kappa \lambda} \df{\delta p_\lambda}{\delta s} = S^{\kappa \lambda} \bigg[ F_\lambda - \df{1}{2}S^{\sigma \mu}R_{\sigma \mu \nu \lambda} t^\nu \bigg] - \df{1}{2M^2} S^{\sigma \mu} S^{\kappa [\lambda} S^{\nu] \tau} R_{\sigma \mu \nu \lambda} \df{\delta p_\tau}{\delta s},\label{ehbe}\ee
where we have used the symmetries of the Riemann tensor to antisymmetrize $\lambda$ and $\nu$ with no cost. In order to simplify this further, we will use the fact that
\be S^{[\kappa \lambda} S^{\nu]\tau} = 0 \label{kilsart} \ee identically due to antisymmetry of $S^{\kappa \lambda}$ and (\ref{cmsr}). The simplest way to see this is to choose a frame where $n_{\kappa}$ takes the form $n_{\kappa}= (1,0,0,0)$ (since it is a unit timelike vector). With this choice, the only nonvanishing components of $S^{\kappa \lambda}$ are $S^{ij}$ where $i,j=1,2,3$, and an explicit calculation, with using the antisymmetry of $S^{\kappa \lambda}$ gives (\ref{kilsart}); which, in turn implies \cite{dixonIV}
\be S^{\kappa [\lambda} S^{\nu] \tau} = \df{1}{2}S^{\nu \lambda} S^{\kappa \tau}. \ee
Then, (\ref{ehbe}) can be written in the form
\be S^{\kappa \lambda} \df{\delta p_\lambda}{\delta s} \bigg[ 1+ \df{1}{4M^2} S^{\sigma \mu} S^{\nu \lambda} R_{\sigma \mu \nu \lambda}\bigg] = S^{\kappa \lambda} \bigg[ F_\lambda -\df{1}{2} S^{\sigma \mu} R_{\sigma \mu \nu \lambda} t^\nu \bigg].\ee

Finally, using (\ref{t}), we have
\be  -[t^\kappa -v^\kappa]\bigg[ M^2 + \df{1}{4} S^{\sigma \mu} S^{\nu \lambda} R_{\sigma \mu \nu \lambda}\bigg] = S^{\kappa \lambda} \bigg[ F_\lambda - \df{1}{2} S^{\sigma \mu} R_{\sigma \mu \nu \lambda} t^\nu \bigg] \label{zdotson}.\ee
This is the desired equation, which gives the velocity of the center of mass $v^{\kappa}$ in terms of other quantities.
In order to check that $v^\kappa$ obtained from this equation satisfies (\ref{cmnorm}), contract with $n^\kappa$. The right hand side vanishes by the center of mass condition. Then, we are left with
\be n_\kappa t^\kappa = n_\kappa v^\kappa .\ee
$t^\kappa$ was defined to be proportional to $n^\kappa + L^{\kappa \lambda} n_\lambda$, so that $n_\kappa t^\kappa = 1$; so the center of mass velocity found above satisfies
\be n_\kappa v^\kappa = 1.\ee

In  a maximally symmetric spacetime, there exists 10 linearly independent Killing vectors so that $F^\kappa = 0 = L^{\kappa \lambda}$ by (\ref{cons1}). If $L^{\kappa \lambda} = 0$, then
\be t^\kappa \sim n^\kappa \label{tpn}.\ee Also, remembering that in such a space, the Riemann tensor can be written as \cite{wald}
\be R_{\kappa \lambda \mu \nu} = 2K g_{\kappa [ \mu} g_{\nu ] \lambda}, \ee  where $K$ is a constant, the left hand side of (\ref{zdotson}) vanishes by (\ref{cmgr}), resulting in
\be v^\kappa = n^\kappa. \label{cmads}\ee
Vanishing of the force already shows that
\be \df{\delta p^\kappa}{\delta s} = 0. \label{pkort}\ee
And using (\ref{cmads}) in (\ref{torkcm}), when $L^{\kappa \lambda}$ vanishes, leads to
\be \df{\delta S^{\kappa \lambda}}{\delta s} = 0.\ee
(\ref{pkort}) also implies
\be \df{\delta M}{\delta s} = 0 .\ee
So that in a maximally symmetric spacetime, the center of mass worldline is a geodesic, angular momentum is parallelly transported along it and $M$ is a constant \cite{dixonIV, de felice, ehlers}.

\subsection{Energy Momentum Skeleton}
Having defined all the necessary concepts in order to define the motion of a body in General Relativity, the next step is to find an explicit expression of force and torque appearing in (\ref{forcegr}) and (\ref{torkcm}). This information is hidden in (\ref{eom}), but there is no straightforward way to obtain it, since that equation contains much more than the information of force and torque. In turn, (\ref{ee}) contains (\ref{eom}), but also some other information about the body and the field. The reason for this is the coupling of the field and the geometry, that is the impossibility of separating matter and field variables. In this section, the aim is to find an expression for the case when one extracts the information of (\ref{eom}) from that of (\ref{ee}). In order to do this, we will write down an expression of the energy momentum tensor, in terms of the momenta (\ref{mom}), (\ref{angmom}) and a new quantity, $\hat{T}^{\kappa \lambda}(z(s), X)$ called the ``Energy Momentum Skeleton", which was first coined by Myron Mathisson \cite{mat}. The name suggests  that this skeleton will have all the necessary information about the body. Mathematically, it will be a tensor valued distribution on the tangent bundle, nonzero only on the tangent spaces of each point of the center of mass worldline, and have a compact support on the $n_\kappa X^\kappa$ hypersurface, where $X^\kappa$ is the position vector in the tangent space of $z$ \cite{dixonIV, schattnerdixon}.
The advantage of this formalism is that instead of dealing with the set of partial differential equations (\ref{eom}), we will be left with two ordinary differential equations (\ref{forcegr}) and (\ref{torkcm}); and we will get explicit forms of force and torque, given an energy momentum tensor \cite{dixonIV, abrahamtez}.

Before starting the discussion summarized above, we will try to obtain the analogs of the idea of moment generating functions discussed in Section (\ref{subsubsec:moments}) in a relativistic theory. After that, we will see how one can the energy momentum skeleton from a given $\emt$; and also how to reconstruct $\emt$ from given $p^\kappa$, $S^{\kappa \lambda}$ and $\skl$. This will be done in the following sections \cite{dixonIV}.

\subsubsection{Moment Generating Functions in a Relativistic Theory}
As stated above, our first aim is to generalize the ideas of Section (\ref{subsubsec:moments}). There, we dealt with a scalar function on $\mathbb{E}^3$, but what we need is the construction for $\emt$ in curved spacetime. There is no straightforward generalization, so we will first work with a scalar function in flat space of Special Relativity, then generalize the results to curved spacetime, and finally obtain the corresponding expression for a rank $2$ tensor field.

We are interested in the energy momentum tensor $\emt$, representing the properties of a finite sized object. The support of such a tensor field will be bounded in spacelike directions, but extend to past and future infinity. Fourier transform of such an object may not be defined in the usual sense because of its support, but one can always define its Fourier transform as a ``generalized function" \cite{gelfand, dixonII}.

We will start with a continuous scalar field $f(x)$ on $\mathcal{E}$, flat space of Special Relativity, whose support extends to past and future infinity, but bounded in spacelike directions. Let $\phi$ be a scalar function of compact support on $\mathcal{E}$, whose Fourier transform is defined as as
\be \phi(x) \rightarrow \tilde{\phi}(k) \equiv \int \phi(x) \exp(i k \cdot x)d^4x.\ee
Then, we can consider $f$ as a functional on the space of all $\phi$, whose value at a particular $\phi$ is
\be \int f \phi \sqrt{-g} d^4x.\ee
If the value of this integral is known for all $\phi$, then $f$ is known. Following this, we will define the Fourier transform of $f$ as a functional on the space of all $\tilde{\phi}$ through Parceval's relation:
\be \int \tilde{f} \tilde{\phi} \sqrt{-g} d^4x = (2\pi)^4 \int f \phi \sqrt{-g} d^4x. \label{cok}\ee
We can decompose this integral by introducing a timelike worldline $L$ with parametric form $z^\alpha (s)$ with $s$ proper time along it, and constructing a family of hypersurfaces $\Sigma(s)$ representing the instantaneous rest space of this observer, constructed through $z(s)$ orthogonal to $L$ for each $s$. This is the worldline along which the multipole expansion will be made. Introduce a vector field $\omega^\alpha$ such that $\omega^\alpha ds$ drags $\Sigma(s)$ to $\Sigma(s+ds)$ for each $s$. Now, if we let
\be \tilde{\phi}(z,k) \equiv \int \phi(x) exp[ik\cdot(x-z)]d^4x = \tilde{\phi}(k) exp(-ik\cdot z). \label{loyloy}\ee
be the Fourier transform of $\phi$ referring $z^\lambda$ as origin; with the help of above construction, (\ref{cok}) can be put in the form \cite{dixonII}
\be \int f \phi \sqrt{-g} d^4x = (2 \pi)^{-4} \int ds \int d^4k \tilde{F}(s,k) \tilde{\phi}(z(s),k). \label{this}\ee
Here $\tilde{F}(s,k)$ is the moment generating function of $f$, analogous of
(\ref{ftild}), defined through the relation
 \be \tilde{F} (s,k) \equiv \sum_{n=0}^\infty \df{-i^n}{n!} k_{\lambda_1} ... k_{\lambda_n} F^{\lambda_1 ... \lambda_n}(s)\label{yettigari}\ee
 with the multipole moments of $f$ defined as integrals over $\Sigma(s)$
 \be F^{\lambda_1 ... \lambda_n}(s) \equiv \int_{\Sigma{s}} r^{\lambda_1} ...
 r^{\lambda_n} f \omega^\alpha d \Sigma_{\alpha}, \quad n \geq 0, \label{yeterulan}\ee
 with $r^{\lambda} \equiv x^\lambda - z^\lambda$.
  (\ref{this}) is the equation
that we will generalize to curved spacetime, but first we should interpret it
properly (We will not give the explicit forms of the analogs of (\ref{yettigari})-(\ref{yeterulan}) in curved spacetime here since they are not important for the rest of this thesis. Interested reader may check \cite{dixonII}). Now, we have taken $z^\lambda$ as origin, so its natural to consider the dot
product in (\ref{loyloy}) as taken between tangent vectors at $z$, so that $k^\lambda$
and $(x-z)^\lambda$ are elements of tangent space $T_z$ at $z$. $(x-z)^\lambda$ is the
element of $T_z$, which is mapped into $x$ by the exponential map $Exp_z$ at $z$. With
this identification, the integral can be considered to be taken over $T_z$, if we
identify $\phi$ on $\mathcal{E}$ with $\phi \circ Exp_z$ on $T_z$. Now, if $z$ is
varied over $\mathcal{E}$, $\tilde{\phi}(z,k)$ will be a scalar field in the tangent
bundle $T(\mathcal{E})$ of $\mathcal{E}$. So, with this construction, we can say that
the Fourier transform relates scalar functions on  $T(\mathcal{E})$ rather than
$\mathcal{E}$.

In a curved spacetime $\mathcal{M}$, we will define the Fourier transform of a function $\Phi$ on $T(\mathcal{M})$ of compact support on $T_z(\mathcal{M})$ for each $z$ as
\be \tilde{\Phi}(z,k) \equiv \int_{T_z} \Phi(z,X) exp (i k \cdot X) DX \ee
with $z \in \mathcal{M}$, $k,X  \in T_z (\mathcal{M})$ and $DX$ is the volume element on $T_z(\mathcal{M})$. Then, the Fourier transform of $f$ is defined as,
\be \int f \phi \sqrt{-g} d^4x = (2\pi)^{-4} \int ds \int Dk \tilde{F}(s,k) \tilde{\Phi}(z(s),k)\ee
Here, there are two different test functions; $\phi$ on $\mathcal{M}$ and $\Phi$ on $T(\mathcal{M})$; which are related by the exponential map of $T(\mathcal{M})$ on $\mathcal{M}$ as $\Phi \equiv \phi \circ Exp$ \cite{dixonII}.

Above ideas can be extended to tensor fields, and particularly to a tensor with contravariant rank $2$ (such as $\emt$). One way to do this in flat space would be considering each component of the tensor field as a scalar function while calculating analogs of (\ref{yeterulan}), but integrating a component of a tensor in curved spacetime will not be covariant, so we need an alternative approach. One was is to introduce a propogator $K$ that can be used to transport the tensor $T^{\alpha \beta}(x)$ to a central world line $z$ before the integration to obtain the multipole moments. If this way is followed, the resulting equation is \cite{dixonII}
\be \int \emt(x) \phi_{\alpha \beta}(x) \sqrt{-g} d^4x = (2 \pi)^{-4} \int ds \int DX \hat{T}^{\kappa \lambda}(s,X) \tilde{\Phi}_{\kappa \lambda}(z(s),X).\ee
Here, $\Phi_{\kappa \lambda}$ is a $C^\infty$ map of compact support of $T(\mathcal{M})$ into the tensor bundle $T_2(\mathcal{M})$, and related to $\phi_{\alpha \beta}$ on $\mathcal{M}$ via some propagator $Z$, which is the inverse of above mention propogator $K$, as
\be \Phi_{\kappa \lambda} (z, X) = {Z^\alpha}_{\lambda}{Z^\beta}_{\kappa} \phi_{\alpha \beta}(x)\ee
where $x = Exp_z X$. So, in order to evaluate the result, one needs the explicit form of this propagator. There are some options for this choice but a natural one would be choosing ${H^{\alpha}}_{\kappa}$ of (\ref{hktanim}) as the propagator such that in Riemann normal coordinates with pole at $z$ (where ${H^{\alpha}}_{\kappa}$ becomes ${\delta^\alpha}_\kappa$), the components of $\phi$ and $\Phi$ would match. Unfortunately, this choice is not useful \cite{dixonIII, abrahamtez}. The correct form requires the introduction of an auxillary field $\Lambda$ satisfying certain conditions which will be discussed in detail in the following section. The resulting equation is \cite{dixonIV, dixonIII}
\be \int \emt(x) \phi_{\alpha \beta}(x) \sqrt{-g}d^4x = \int ds \int \hat{T}^{\kappa \lambda} \bigg( \Phi_{\kappa \lambda} + {G^\mu}_{\kappa \lambda} \Lambda _\mu \bigg)DX. \label{lololo}\ee
This is the equation which will be examined in the following sections, where the terms appearing in (\ref{lololo}) will be explained in detail.

\subsubsection{Constraints, Relation to the Energy Momentum Tensor}

In this section, we will directly write down the equation relating $\emt$ to $p^\kappa, S^{\kappa \lambda}$ and $\skl$ and the constraint equations on $\skl$, as they appear in \cite{dixonIV, dixonIII}, and give the proof of the existence of such a $\skl$, without discussing the reason of the choice. Details of this construction is beyond the scope of this thesis and can be found in \cite{dixonIII}. We will start by separating monopole and dipole contributions to $\skl$ in (\ref{lololo}), which is actually separating the moments whose time evolution is determined by (\ref{eom}) (see Section \ref{subsec:Relativistic Multipole Moments} and (\ref{extskl})). Then, (\ref{lololo}) takes the form

\be \int \emt \phi_{\alpha \beta} \keg d^4x = \int ds \bigg[ p^\kappa v^\kappa \phi_{\kappa \lambda} + S^{\kappa \lambda} v^\mu \nabla_\kappa \phi_{\lambda \mu} + \int \skl \bigg(\Phi_{\kappa \lambda} + {G^\mu}_{\kappa \lambda} \Lambda_\mu \bigg)DX\bigg],\label{emtskl}\ee
for all symmetric $\phi_{\alpha \beta}$ of compact support. Here $DX$ is the invariant volume element on the tangent space of $z$ and
\be \Phi_{\kappa \lambda}(z,X) \equiv {H^\alpha}_\kappa {H^\beta}_\lambda \phi_{\alpha \beta}(x), \label{Phi}\ee
\be {G^\mu}_{\kappa \lambda}(z,X) \equiv {H^\alpha}_\kappa {H^\beta}_\lambda {\sigma^\mu}_{\alpha \beta},\label{Gmunu}\ee
and
\be \Lambda_\mu (z,X)\equiv {H^\alpha}_\kappa \lambda_\alpha(z,x),\label{Lambda}\ee
such that $\lambda_\alpha$ satisfies the inhomogeneous generalization of the Jacobi equation (See also (\ref{egd}))
\be \df{\delta^2 \lambda_\alpha}{\delta u^2} + {R_{\alpha \beta\gamma}}^\delta \dot{x}^\beta \dot{x}^{\gamma} \lambda_\delta = \dot{x}^\beta \dot{x}^\gamma \nabla_{\{\beta}\phi_{\alpha \gamma \}},\label{lamjac}\ee
for all geodesics $x(u)$ through $z$ with $u$ an affine parameter along them with curly bracket notation defined only for three indices expressing
 \be t_{\{\alpha \beta \gamma\}} = t_{\alpha \beta \gamma} - t_{\beta \gamma \alpha} + t_{\gamma \alpha \beta },\label{curly}\ee and such that $\lambda_\alpha$ satisfies the following boundary conditions:
\be \lim_{x \rightarrow z} \lambda_\alpha \rightarrow 0\quad,\quad \lim_{x \rightarrow z} (\nabla_\alpha \lambda_\beta - \phi_{\alpha \beta})  \rightarrow 0.\label{lamboun}\ee
In this expression, it should be understood that the derivatives are taken before the limits. In all the equations above, we used indices as defined in Chapter \ref{sec:notation} \cite{dixonIV}
\subsubsection{Existence Proof}

As a first step in the proof, we will show that for all symmetric $\phi_{\alpha \beta}(x)$ of compact support, there exists a unique symmetric $e_{\alpha \beta}(x)$ such that
\be e_{\alpha \beta}(x) \sigma^\beta(z,x) = 0 \label{cond1},\ee
and
\be \phi_{\alpha \beta} = e_{\alpha \beta} + \nabla_{(\alpha} \lambda_{\beta)} \label{cond2},\ee
such that
\be \lambda_\mu(z) = 0 = \nabla_{[\mu}\lambda_{\nu]}(z) \label{lamin}.\ee

In order to prove this, we first prove the following lemma \cite{dixonIV, dixonI, schattnerdixon}:\\
\textit{Lemma 1}:
The condition (\ref{cond1}) is equivalent to the following
\be \sigma^\beta \sigma^\gamma \nabla_{\{ \beta}e_{\alpha \gamma \}} = 0\label{lemma11},\ee
and
\be e_{\kappa \lambda}(z) = 0\label{lemma12}.\ee

\textit{Proof}:
First we will show how to get the latter condition from the former one. Taking the covariant derivative of (\ref{cond1}) and expanding it, we have
\be {\sigma^\alpha}_\gamma e_{\alpha \beta} + \sigma^\alpha \nabla_\gamma e_{\alpha \beta} = 0\label{proof11}.\ee
Multiplying this with $\sigma^\beta$ ($\sigma^\gamma$) gives, using (\ref{cond1}) and (\ref{wfss})
\be \sigma^\beta \sigma^\alpha \nabla_{\gamma} e_{\alpha \beta} = 0\quad,\quad\quad \sigma^\gamma \sigma^\alpha \nabla_{\gamma} e_{\alpha \beta} = 0.\ee
Using the symmetry, these two expressions together are equivalent to (\ref{lemma11}).
In order to find ({\ref{lemma12}}), take the limit of (\ref{proof11}), as $x \rightarrow z$. In this limit, $\sigma^\alpha_\gamma \rightarrow \delta^\alpha_\gamma$, and $\sigma^\alpha = 0$. Then we are left with
\be e_{\kappa \lambda}(z) = 0.\ee
This completes the second part of the proof. Now, we will show that (\ref{lemma11}) and (\ref{lemma12}) implies (\ref{cond1}).
Suppose that the conditions (\ref{lemma11}) and (\ref{lemma12}) hold. Contracting (\ref{lemma11}) with $\sigma^\alpha$ gives
\be \sigma^\alpha \sigma^\beta \sigma^\gamma \nabla_{\{ \beta}e_{\alpha \gamma \}} = 0.\ee
Using (\ref{wfxx}) this can be written as
\be \df{\delta}{\delta u} \bigg( \dot{x}^\alpha \dot{x}^\beta e_{\alpha \beta}\bigg) = 0,\ee
since $\sigma^\beta \nabla_\beta = u\dot{x}^\beta \nabla_\beta = u \df{\delta}{\delta u}$(which differentiates along geodesics $x(u)$ through $z(s)$).
This can be integrated by using (\ref{lemma12}) as the initial data. Then,
\be \dot{x}^\alpha \dot{x}^\beta e_{\alpha \beta} = \sigma^\alpha \sigma^\beta e_{\alpha \beta} = 0.\ee
Taking the covariant derivative, $\nabla_\gamma$, of this, and rearranging the terms gives
\be 2\sigma^\beta_\alpha \sigma^\gamma e_{\beta \gamma} + \sigma^\beta \sigma^\gamma \nabla_\alpha e_{\beta \gamma} = 0.\ee
Using this in (\ref{lemma11}) gives:
\be \sigma^\beta_\alpha \sigma^\gamma e_{\beta \gamma} = - \sigma^\beta \sigma^\gamma \nabla_\beta e_{\alpha \gamma}\label{lemma113},\ee
which can be put in a more convenient form by noticing that the term on the left hand side can be rewritten by using
\be  \df{\delta}{\delta u}\bigg(\df{1}{u}  (e_{\alpha \beta} \dot{x}^\beta)\bigg) = \dot{x}^\gamma (\nabla_\gamma e_{\alpha \beta})\dot{x}^\beta = \df{1}{u^2}\sigma^\beta \sigma^\gamma \nabla_\gamma e_{\alpha \beta}.\ee
So (\ref{lemma113}) can be written as
\be \df{\delta}{\delta u} \bigg(\df{1}{u}e_{\alpha \beta}\sigma^\beta    \bigg) + \df{1}{u^2} {\sigma^\beta}_\alpha \sigma^\gamma e_{\beta \gamma} = 0.\ee
Writing this as a linear equation for $e_{\alpha \beta} \sigma^\beta$, by taking the $u$ in the first term out of the derivative, one obtains
\be u\df{\delta}{\delta u}(e_{\alpha \beta} \sigma^\beta) + ({\sigma^\gamma}_\alpha - \delta^\gamma_\alpha)e_{\gamma \beta}\sigma^\beta = 0    , \ee
which has the unique solution $\sigma^\beta e_{\alpha \beta} = 0$, by the use of (\ref{lemma12}) as the initial condition. This is what we were looking for, that is (\ref{cond2}). This completes the proof.

By directly using (\ref{cond2}) in (\ref{lemma12}), it is straightforward to show that this implies (\ref{lamjac}) by using
\be [\nabla_\alpha, \nabla_\beta] \lambda_\alpha = - {R_{\alpha \beta \gamma}}^\delta \lambda_\delta . \label{partrans}\ee

Now, note that by (\ref{cond1}), (\ref{lamin}) and (\ref{lemma12}) together imply
\be \lambda_\kappa (z) = 0\quad,\quad\quad \nabla_{\kappa}\lambda_{\mu}(z)= \phi_{\kappa \mu}(z).\label{lamjacin1}\ee
This can be used as initial conditions for (\ref{lamjac}), by rewriting them as
\be \lambda_\kappa(x(0)) = 0,\ee
and
\be \df{\delta}{\delta u} \lambda_{\kappa} (x(0)) = \dot{x}^\mu(0)\phi_{\kappa \mu}(0).\ee
Then one can integrate (\ref{lamjac}) along $x(u)$ to get the $\lambda_\alpha$ that satisfies (\ref{cond1}) and (\ref{lamin}) \cite{dixonIII, schattnerdixon}.

Up to now, calculations were made for an arbitrary $z$. Now, we can extend the result by treating $\lambda_\alpha$ and $e_{\alpha \beta}$ as two point tensor that have scalar character at $z(s)$ \cite{dixonIV, schattnerdixon}. With this construction, (\ref{cond2}) can be written as
\be \phi_{\alpha \beta}(x) = e_{\alpha \beta}(x,z) + \nabla_{(\alpha} \lambda_{\beta)}(x,z) \label{cond2y}.\ee
In order to extract the $s$ dependence of $\lambda_\alpha$, take the derivative with respect to $s$ to get
\be \df{\partial e_{\alpha \beta}}{\partial s} + \nabla_{(\alpha} \df{\partial}{\partial s} \lambda_{\beta )} = 0.\ee
Now, comparing with (\ref{cond2}) shows that this can be obtained from (\ref{cond2}) if we set $e_{\alpha \beta} \rightarrow  0$, $\phi_{\alpha \beta} \rightarrow -\df{\partial e_{\alpha \beta}}{\partial s}$ and $\lambda_\alpha \rightarrow \df{\partial \lambda_\alpha}{\partial s}$.
Note that this construction identically satisfies (\ref{cond1}). Then, by \textit{Lemma 1}, (\ref{lamjac}) is also satisfied in the following form
\be \df{\delta^2}{\delta u^2} \bigg( \df{\partial \lambda_\alpha}{\partial s}\bigg) + {R_{\alpha \beta\gamma}}^\delta \dot{x}^\beta \dot{x}^\gamma \bigg( \df{\partial \lambda_\alpha}{\partial s}\bigg) = \dot{x}^\beta \dot{x}^\gamma \nabla_{\{\beta} \bigg( -\df{\partial}{\partial s} e_{\alpha \gamma \}}\bigg)\label{lamjacs}.\ee
By (\ref{lemma11}), we have
\be \df{\partial}{\partial s} \bigg( \sigma^\beta \sigma^\gamma \nabla_{\{\beta} e_{\alpha \gamma \}}\bigg)=0.\ee
Then
\be \bigg( \df{\partial}{\partial s}\sigma^\beta\bigg)\sigma^\gamma \nabla_{\{\beta} e_{\alpha \gamma \}} + \sigma^\beta \bigg( \df{\partial}{\partial s}\sigma^\gamma \bigg)\nabla_{\{\beta} e_{\alpha \gamma \}}  + \sigma^\beta \sigma^\gamma \nabla_{\{\beta} \df{\partial}{\partial s}e_{\alpha \gamma \}}=0.\ee
Using $\df{\partial}{\partial s} = \df{dz^\kappa}{ds}\df{\partial}{\partial z^\kappa} = v^\kappa\df{\partial}{\partial z^\kappa} $, this can be written as
\be v^\kappa {\sigma^\beta}_\kappa \sigma^\gamma  \nabla_{\{\beta} e_{\alpha \gamma \}} + \sigma^\beta v^\kappa{\sigma^\gamma}_\kappa  \nabla_{\{\beta} e_{\alpha \gamma \}} +\nabla_{\{\beta} \df{\partial}{\partial s}e_{\alpha \gamma \}}=0 .\ee
By relabeling dummy indices in the first two terms, it can be seen that they are equivalent. Then (\ref{lamjacs}) can be written as, by using (\ref{wfxx})
\be \df{\delta^2}{\delta u^2} \bigg( \df{\partial \lambda_\alpha}{\partial s}\bigg) + {R_{\alpha \beta\gamma}}^\delta \dot{x}^\beta \dot{x}^\gamma \bigg( \df{\partial \lambda_\alpha}{\partial s}\bigg) = 2 \sigma^\beta {\sigma^\gamma}_\kappa v^\kappa  \nabla_{\{\beta} \bigg( -\df{\partial}{\partial s} e_{\alpha \gamma \}}\bigg).\ee
In order to integrate this and to find a solution for $\df{\delta \lambda_\alpha}{\delta s}$, we need the initial conditions that correspond to (\ref{lamjacin1}). This can be found by differentiating them with respect to $s$ and taking the coincidence limit, $x \rightarrow z$ afterwards. The resulting initial conditions, obtained by using (\ref{syngesrule}) are \cite{dixonIV, schattnerjacobi}
\be \df{\partial \lambda_\mu}{\partial s} = -v^\nu \phi_{\mu \nu}\quad,\quad \quad
 \nabla_\mu \bigg( \df{\partial \lambda_\nu}{\partial s}\bigg) = -2v^\rho
  \nabla_{[\mu}\phi_{\nu] \rho}.\ee
Now, let us assume that we have the solution of (\ref{lamjacs}) with the initial conditions
\be \df{\partial \lambda_\mu}{\partial s} = 0\quad,\quad\quad\nabla_\mu \bigg( \df{\partial \lambda_\nu}{\partial s}\bigg) = 0.\ee
Let this solution be $\psi_\alpha(s,x)$. This solution is determined only by $e_{\alpha \beta}$ and its exact form can be found in \cite{schattnerdixon}. Then, the solution of (\ref{lamjacs}) can be obtained by subtracting the solution of the homogeneous part. This part can be written in terms of its initial values by (\ref{jacgensol}), the corresponding solution to the inhomogeneous Jacobi equation is of the form \cite{dixonIV, schattnerdixon}
\be \df{\partial \lambda_\alpha}{\partial s} = \psi_\alpha - {K_\alpha}^\mu v^\nu \phi_{\mu \nu} - 2 {H_\alpha}^{[\mu}\sigma^{\nu ]} v^\rho \nabla_\mu \phi_{\nu \rho}.\label{psitamcoz}\ee

Before introducing the energy momentum tensor, let us discuss one further remark. Let us discuss the results of \textit{Lemma 1} in flat space (such as the tangent space at a point). Let $X^\alpha$ be the coordinates  on $\mathbb{R}^4$ and $\Psi_{\alpha \beta}(X)$ be a symmetric tensor field. Then one can prove that there exists a tensor field \cite{schattnerdixon}
\be H_{\alpha \beta \gamma \delta}(X) = H_{[\gamma \delta][\alpha \beta]}\label{hsym},\ee with
\be \Psi_{\alpha\beta}(X) = X^\gamma X^\delta H_{\alpha \gamma \beta \delta}\label{psih}, \ee
for all $X$, from \textit{Lemma 1}.
It can be shown that with such an $H$, $\Psi$ satisfies (\ref{cond1}) and (\ref{cond2})
 by its symmetries. In order to shown the other way of the proof, take the
  derivative, $\partial_\gamma \equiv \df{\partial}{\partial X^\gamma}$ of
   (\ref{cond1}) after the replacements $\sigma^\mu \rightarrow X^\mu$,$e_{\alpha \beta} \rightarrow \Psi_{\alpha \beta}$. We obtain
\be \Psi_{\gamma \beta} + X^\alpha \partial_\gamma \Psi_{\alpha \beta} = 0\label{psi1}.\ee
Since $\Psi_{\alpha \beta}$ is symmetric, we have
\be X^\alpha \partial_{[\gamma}\Psi_{\beta]\alpha} = 0\label{psisym}.\ee
Operating on this by $\partial_\delta$, one finds
\be \partial_\delta \bigg( X^\alpha \partial_{[\gamma}\Psi_{\beta]}\alpha	\bigg) = 0.\ee
Distributing the derivative and contracting with $X^\gamma$, one gets
\be X^\gamma \partial_{[\alpha}\Psi_{\beta]\delta} + X^\gamma X^\alpha \partial_{\delta [\gamma}\Psi_{\beta]}\alpha  = 0 \label{xasym}.\ee
Now introduce $f_{\alpha \beta}(u)$ such that
\be f_{\alpha \beta}(u) \equiv u\Psi_{\alpha \beta}(uX),\ee
with
\be f_{\alpha \beta}(0) = 0 \quad,\quad\quad f_{\alpha \beta}(1) = \Psi_{\alpha \beta}(X).\ee
Then
\be \df{\delta}{\delta u}f_{\alpha \beta}(u) = \Psi_{\alpha \beta}(uX) + u\df{\partial}{\partial u} \Psi_{\alpha \beta}(uX)\label{fbirtur}.\ee
Using
\be \df{\partial}{\partial u} = \df{\partial (uX^\alpha)}{\partial u}\df{\partial}{\partial(uX^\alpha)} = X^\alpha\df{\partial}{\partial(uX^\alpha)}, \label{uxtur}\ee
(\ref{fbirtur}) becomes
\be \df{\partial}{\partial u}f_{\alpha \beta}(u) = \Psi_{\alpha \beta}(uX) + u X^\gamma \partial_\gamma \Psi_{\alpha \beta}.\ee
Using these results, let us set $X \rightarrow uX$ in (\ref{psi1})
\be \Psi_{\alpha \gamma}(uX) + u X^\delta \partial_\gamma \Psi_{\alpha \delta}(uX) = 0.\ee
By following the remarks made after (\ref{psi1}), we have
\be \df{\partial}{\partial u}f_{\alpha \beta}(u) = 2u X^\gamma \partial_{[\gamma}\Psi_{\alpha ] \beta}(uX) \label{f2tur}.\ee
Note that for $u=0$, $\df{\partial f_{\alpha\beta}}{\partial u} = 0$.
Taking another derivative of (\ref{f2tur}) with respect to $u$, and by using (\ref{uxtur}) again
\be \df{\partial^2}{\partial u^2} f_{\alpha \beta}(u) = 2X^\gamma \partial_{[\gamma}\Psi_{\alpha ] \beta} + 2u X^\gamma X^\rho \partial_{\rho[\gamma}\Psi_{\alpha ] \beta}\label{f2tur1} .\ee
Let $X \rightarrow uX$ in (\ref{xasym}) to get
\be uX^\gamma \partial_{[\gamma}\Psi_{\alpha ] \beta} = u^2 X^\gamma X^\delta \partial_{\beta[\alpha}\Psi_{\gamma]\delta}.\ee
Then (\ref{f2tur1}) can be written as
\be \df{\partial^2}{\partial u^2}f_{\alpha\beta}(u) = 4uX^\gamma X^\delta \partial_{[\alpha [\beta}\Psi_{\gamma]\delta]}(uX),\ee
where the antisymmetrization is over the pairs $(\alpha, \gamma)$ and $(\beta, \delta)$ on the left hand side.
Now, $f_{\alpha \beta}(1)$ can be written as
\be f_{\alpha \beta}(1) = f_{\alpha \beta}(0) + \df{\partial}{\partial u}f_{\alpha \beta}(u) + \int_{0}^1 (1-u)\df{\partial}{\partial u} f_{\alpha \beta}(u) du,\ee
as can be seen by carrying out the integral on the right hand side. Using these results for the first two terms on the right hand side and using $f_{\alpha \beta} (1) = \Psi_{\alpha \beta}(X)$, we obtain
\be \Psi_{\alpha \beta}(X) = f_{\alpha \beta}(1) = X^{\gamma}X^\delta H_{\alpha \gamma \beta \delta }(X), \ee
with
\be H_{\alpha \beta \gamma\delta}(X) = 4\int_{0}^1 u(1-u) \partial_{[\alpha [\gamma}\Psi_{\beta]\delta]}(uX)du.\label{htanim}\ee
This completes the proof and gives the exact form of $H$ \cite{schattnerdixon}.

\subsubsection{Construction of the Energy Momentum Skeleton}

Now that we've finished the construction of mathematical preliminaries, we will obtain (\ref{emtskl}). In this construction, we need a future pointing timelike unit vector $n^\kappa$ and the hypersurfaces of integration constructed as explained in the paragraph preceeding (\ref{nsig}). In addition to these, we will define a scalar field $\tau(x) = s$ for all $x$ on $\Sigma(s)$ and a vector field $\omega^\alpha(x)$ such that it maps $\Sigma(s) \rightarrow \Sigma(s +ds)$, with $\omega^\alpha \partial_\alpha \tau = 1$. With these at hand, consider the right hand side of (\ref{emtskl}). In the view of \textit{Lemma 1}, it can be written as \cite{dixonIV}
\be \int \emt \phi_{\alpha \beta} \keg d^4x = \int \emt \bigg(e_{\alpha \beta} + \nabla_{(\alpha}\lambda_{\beta)} \bigg)\keg d^4x,\ee
which can, by using the definition of $\omega_\alpha$, be written as \cite{dixonIV, dixon64, dixonIII}
\be \int \emt \phi_{\alpha \beta} \keg d^4x = \int ds \int_{\Sigma(s)} \bigg( \emt e_{\alpha \beta} \omega^\alpha - T^{\beta \gamma} \df{\partial \lambda_\beta}{\partial s}\bigg) d\Sigma_\gamma .\ee
Using (\ref{psitamcoz}), and plugging in the definitions (\ref{mom}) and (\ref{angmom}):
\be\int \emt \phi_{\alpha \beta} \keg d^4x = \int ds \bigg[ p^\mu v^\nu \phi_{\mu \nu} + S^{\mu \nu} v^\rho \nabla_\mu \phi_{\nu \rho} + \int_{\Sigma(s)}\bigg(\emt \phi_{\alpha \beta}\omega^\gamma - T^{\beta \gamma} \psi_\beta \bigg)d\Sigma_\gamma  \bigg]. \ee
Comparing this with (\ref{emtskl}) shows that the last term (integral over $d\Sigma$) is identified with the energy momentum skeleton. It is a scalar valued functional of $e_{\alpha \beta}$. Let
\be t_s[e_{\alpha \beta}] \equiv \int_{\Sigma(s)}\bigg(\emt \phi_{\alpha \beta}\omega^\gamma - T^{\beta \gamma} \psi_\beta \bigg)d\Sigma_\gamma \label{ts}.\ee
When $z(s)$ is on the center of mass worldline, we will set
\be \int_{M_z} \skl \Psi_{\kappa \lambda} DX = t_s[e_{\alpha \beta}]. \label{sklts}\ee
Here $DX$ is the volume element on the tangent space $M_z$ to the manifold at $z$, and
\be e_{\alpha \beta}(z,x) =  {\sigma^\kappa}_\alpha {\sigma^\lambda}_\beta E_{\kappa \lambda}(z,X)\label{eE},\ee
where $E_{\kappa \lambda}(z,x)$ is a tensor field defined in the same way as $\Psi$ was in \ref{psih}, i.e. it is equal to $\Psi$ when results of $Lemma 1$ holds in flat space.

Now, we will examine the properties of $\skl$ defined in this way \cite{dixonIV}. In order to get the first important property, take
\be \Psi_{\kappa \lambda} = \nabla_{\ast (\kappa} \Psi_{\lambda)},\ee
where $\nabla_{\ast \kappa}$ is as defined in (\ref{vercovder}). With this choice, $H_{\alpha \beta \gamma\delta }$ vanishes by (\ref{htanim}), so (\ref{ts}) also vanishes by construction. So, one can integrate by parts the left hand side of (\ref{sklts}) and use the compact support of $\skl$ to get
\be \int \bigg( \nabla_{\ast \lambda} \skl \bigg) \Psi_{\kappa} DX = 0 \label{sklver}\ee
for arbitrary $\Psi_\kappa(z,X)$. Then we have
\be \nabla_{\ast \kappa} \skl = 0. \label{sklcons}\ee
Now, we will derive another property of $\skl$ which will restrict the singularity that must be present to concentrate the distribution on the hypersurface $n_\kappa X^\kappa=0$. For this, define a new tensor field $K_{\alpha\beta\gamma\delta}$, with the same symmetry properties of (\ref{hsym}) and set
\be \Psi_{\kappa \lambda} = (n_\tau X^\tau) X^{\mu}X^\nu K_{\kappa \mu \lambda \nu}.\ee
With such a $\Psi$, one has $X^\lambda \Psi_{\kappa \lambda} = 0$, so $E_{\kappa \lambda} = \Psi_{\kappa \lambda}$ by \textit{Lemma 1}. By using this result with (\ref{nsig}) and (\ref{wfss}) on (\ref{eE}), $t_s[e_{\alpha\beta}]$ vanishes. Then, from (\ref{sklts}), we have
\be \int_{M_z} \skl (n_\tau X^\tau) X^\mu X^\nu K_{\kappa \mu \lambda \nu} DX = 0. \ee
By the arbitrariness of $K$, we have
\be  \skl (n_\tau X^\tau) X^\mu X^\nu = 0.\label{jdekul}\ee
Now finally, we are going to show how to get the form (\ref{emtskl}) as it was first introduced \cite{dixonIV}. For this, let us rewrite the equation explicitly
\be \int \emt e_{\alpha\beta} \keg d^4x = \int ds \bigg( p^\kappa v^\lambda \phi_{\kappa \lambda} + S^{\kappa \lambda} v^\mu \nabla_\kappa \phi_{\lambda \mu} + \int \skl \Psi_{\kappa \lambda} DX\bigg) .\label{bitmedi2}\ee
Now rewrite (\ref{eE}) by using (\ref{hktanim})
\be E_{\kappa \lambda} = {H^\alpha}_\kappa {H^\beta}_\lambda e_{\alpha \beta}.\ee
Using (\ref{cond2})
\be E_{\kappa \lambda} = {H^\alpha}_\kappa {H^\beta}_\lambda \phi_{\alpha \beta} - {H^\alpha}_{\kappa} {H^\beta}_{\lambda} \nabla_{(\alpha } \lambda_{\beta)}.\label{bitmedi}\ee
Here the first term gives $\Phi_{\kappa \lambda}$, as defined in  (\ref{Phi}). In order to evaluate the second term, first rewrite it as:
\be {H^\alpha}_\kappa {H^\beta}_\lambda = {H^\alpha}_\kappa \nabla_{(\alpha}({H^\beta}_{|\lambda |} \lambda_{\beta)}) - {H^\alpha}_{\kappa } (\nabla_{(\alpha}{H^\beta}_{|\lambda |}) \lambda_{\beta)} \label{ucsigma}.\ee
Now, by (\ref{hktanim})
\be {\sigma^\lambda}_\alpha {H^\beta}_\alpha = - {\delta^\beta}_\gamma.\ee
Taking the covariant derivative of this, one has
\be {\sigma^\lambda}_{\gamma \alpha} {H^\beta}_\lambda + {\sigma^\lambda}_\gamma \nabla_\alpha {H^\beta}_\lambda = 0.\ee
Then, we have
\be \nabla_\alpha {H^\beta}_\tau = {H^\gamma}_\tau {H^\beta}_\lambda {\sigma^\lambda}_{\gamma \alpha},\label{sigmauch}\ee
again by using (\ref{hktanim}). Using (\ref{sigmauch}), the definitions (\ref{Gmunu}) and (\ref{Lambda}), we have, for the second term on the right hand side of (\ref{bitmedi})
\be {H^\alpha}_{(\kappa} \nabla_{|\alpha |} \Lambda_{\lambda)} - {G^\mu}_{\kappa \lambda} \Lambda_{\mu}.\ee
The first term on this expression does not contribute to the integral (\ref{bitmedi2}), since by (\ref{vcdr}), it can be written as
\be \nabla_{\ast (\kappa}\Lambda_{\lambda)}, \label{vanishcov}\ee
whose contribution to the integral vanishes in the view of (\ref{sklver}). So we finally obtain the form given in (\ref{emtskl}).

Now we are in a position to show that, if $\emt$ satisfies (\ref{emtskl}), then it automatically satisfies $\nabla_\alpha \emt = 0$. In order to do this \cite{dixonIV}, choose $\phi_{\alpha \beta}$ such that
\be \phi_{\alpha \beta} = \nabla_{(\alpha}\phi_{\beta )}(x),\label{phi1}\ee
for an arbitrary $\phi_\alpha (x)$ of compact support which vanishes in some neighbourhood of the center of mass worldline. Use this in (\ref{lamjac})
\be  \df{\delta^2 \lambda_\alpha}{\delta u^2} + R_{\alpha \beta\gamma}^\delta \dot{x}^\beta \dot{x}^{\gamma} \lambda_\delta = \dot{x}^\beta \dot{x}^\gamma \nabla_{\{\beta}\nabla_{(\alpha}\phi_{\gamma)\} }.\ee
Expanding the terms, using the geodesic equation and (\ref{partrans}), it can be seen that
\be \lambda_\alpha (z,x) = \phi_\alpha (x)\label{eskilambda}\ee
is a solution, since by the definition of the support of $\phi_\alpha$, the initial conditions (\ref{lamboun}) hold.
Now introduce a new tensor $\Phi_\kappa$ as
\be \Phi_{\kappa} \equiv {H^{\alpha}}_\kappa \phi_\alpha. \ee
Form this definition, let us calculate the (symmetrized) vertical derivative of $\Phi_\kappa$
\be \nabla_{\ast ( \kappa}\Phi_{\lambda )} = \nabla_{\ast ( \kappa} ({H^\alpha}_{\lambda)} \phi_\alpha).\label{phicovt}\ee
Expand the terms to get
\be \nabla_{\ast ( \kappa}\Phi_{\lambda )} = \df{1}{2} \bigg[  \nabla_{\ast \kappa}({H^\alpha}_\lambda \phi_\alpha) + \nabla_{\ast \lambda}({H^\alpha}_\kappa \phi_\alpha)\bigg].\ee
Using (\ref{sigmauch}), this can be written as:
\be \nabla_{\ast ( \kappa}\Phi_{\lambda )} =\df{1}{2} \bigg[\bigg({H^\beta}_\kappa {H^\gamma}_\lambda + {H^\beta}_\kappa {H^\lambda}_\kappa \bigg) \bigg(\phi_\alpha {H^\alpha}_\mu {\sigma^\mu}_{\gamma \beta}  + \nabla_\beta \phi_\alpha \bigg) \bigg].\label{buyukphi}\ee
Using the definitions (\ref{Phi}), (\ref{Gmunu}) and (\ref{Lambda}), (\ref{phicovt}) can be written as
\be \nabla_{\ast(\kappa}\Phi_{\lambda )} = \Phi_{\kappa \lambda} + {G^\mu}_{\kappa \lambda} \Lambda_\mu.\ee
Using this in (\ref{emtskl}); while keeping in mind the comments on (\ref{vanishcov}) and the support of $\phi_\alpha$, the right hand side vanishes. Then integrating the rest by parts, one is left with
\be \int \phi_\alpha \nabla_\beta \emt \keg d^4x = 0 ,\ee
for all $\phi_\alpha$ with described support. Then, by the arbitrariness of $\phi_\alpha$, one obtains (\ref{eom}) \cite{dixonIV}.

\subsection{Gravitational Force and Torque}

In this section, the analogs of (\ref{forcegr}) and (\ref{torkcm}) will be derived. In order to do this, again take $\phi_{\alpha \beta}$ to be of the form (\ref{phi1}), but this time assume $\phi_\alpha$ is nonzero in the neighborhood of the center of mass worldline \cite{dixonIV}. Around this worldline, let
\be \phi_\kappa (z(s)) \equiv A_\kappa (s)\quad,\quad\quad \nabla_{[\kappa} \phi_{\lambda]}(z(s)) \equiv B_{\kappa \lambda}(s).\label{yeniphiini}\ee
With such a choice, we still have $\lambda_\alpha(z,x) =\phi_{\alpha} $ as a solution of (\ref{lamjac}), but its initial conditions must be set by using the method of (\ref{psitamcoz}), in order to get the form (\ref{lamboun}). So, with the help of a Jacobi field $\xi^\alpha$ along a fixed geodesic with
\be \xi_\kappa = A_\kappa, \quad \nabla_\kappa \xi_\lambda = B_{\kappa \lambda},\label{xiinitial}\ee
it is found that
\be \lambda_\alpha (z(s),x) = \phi_\alpha(x) - \xi_\alpha(s,x),\label{yenilambda}\ee
with
\be \xi_\alpha (s,x) \equiv {K_\alpha}^\kappa A_\kappa + {H_\alpha}^\kappa \sigma^\lambda B_{\kappa \lambda},\label{xiyeni}\ee
is the needed solution \cite{dixonIV}.
At this point, we would like to make use of (\ref{buyukphi}), but now $\lambda_\alpha$ is defined as in (\ref{yenilambda}) instead of (\ref{eskilambda}), so (\ref{buyukphi}) must be modified as
\be \nabla_{\ast (\kappa} \Phi_{\lambda)}=\nabla_{\ast (\kappa} \Theta_{\lambda)} + {H^\alpha}_\kappa {H^\beta}_\lambda \nabla_{(\alpha}\xi_{\beta)},\label{yenitheta}\ee
where
\be \Theta_\kappa(s,X) \equiv {H^\alpha}_\kappa \xi_\alpha (s,x).\ee
By checking (\ref{buyukphi}), one can see that $\lambda_\alpha$ only enters the definition through the term ${G^\mu}_{\kappa \lambda} \Lambda_\mu$. Using (\ref{yenilambda}) here will give an extra term of the form
\be {H^\gamma}_\kappa {H^\beta}_\lambda {\sigma^\mu}_{\gamma \beta}{H^\alpha}_\mu \xi_\alpha.\ee
Now we will show that the extra terms in (\ref{yenitheta}) is equivalent to this. Using the given definitions and (\ref{vcdr})
\be -\nabla_{\ast (\kappa} \Theta_{\lambda)} + {H^\alpha}_\kappa {H^\beta}_\lambda \nabla_{(\alpha}\xi_{\beta)} = - {H^\alpha}_\kappa \nabla_\alpha \bigg({H^\gamma}_\lambda \xi_\gamma	\bigg) + {H^\alpha}_\kappa {H^\beta}_\lambda \nabla_{(\alpha} \xi_{\beta)}\ee
Using (\ref{sigmauch}) this can be written as
\be =-{H^\alpha}_\kappa {H^\beta}_\mu {H^\gamma}_\lambda {\sigma^\mu}_{\gamma\alpha}\xi_\beta - {H^\alpha}_\kappa (\nabla_\alpha \xi_\beta) {H^\beta}_\lambda + {H^\alpha}_\kappa {H^\beta}_\lambda \nabla_{(\alpha} \xi_{\beta)}\ee
\be ={H^\gamma}_\kappa {H^\beta}_\lambda {\sigma^\mu}_{\gamma \beta}{H^\alpha}_\mu \xi_\alpha.\ee
At this point, we need another manipulation of ${H^\alpha}_\kappa$. By (\ref{hktanim}), we have \cite{dixonIV}
\be \mathfrak{L}_\xi ({\sigma^\lambda}_\gamma {H^\beta}_\lambda) = \mathfrak{L}_\xi (-{\delta^\beta}_\gamma) = 0.\ee
Also by (\ref{sigmalie})
\be \nabla_\alpha \mathfrak{L}_\xi \sigma^\kappa = \mathfrak{L} ({\sigma^\kappa}_\alpha) = 0,\ee
since $\sigma^\kappa$ is a scalar at the point $x$ so that the Lie derivative and covariant derivative at that point commutes. From these, we have
\be \mathfrak{L}_\xi {H^\beta}_\lambda = 0.\ee
Also since,
\be \nabla_{(\alpha} \xi_{\beta)} = \df{1}{2}\mathfrak{L}_\xi g_{\alpha \beta},\ee
the last term in (\ref{yenitheta}) can be written as
\be {H^\alpha}_\kappa {H^\beta}_\lambda \nabla_{(\alpha} \xi_{\beta )} = \df{1}{2} \mathfrak{L}_\xi G_{\kappa \lambda},\ee
with
\be G_{\kappa \lambda} \equiv {H^\alpha}_\kappa {H^\beta}_\lambda g_{\alpha \beta}.\label{buyukg}\ee
For fixed $z$, in normal coordinates with pole at $z$, we have $\sigma^\kappa (z,x) = - x^\kappa$ for all $x$. Then, from the definitions, ${H^\alpha}_\kappa = \delta^\alpha_\kappa$ so that $G_{\kappa \lambda} = g_{\kappa \lambda}$.
It is time to use these tools in (\ref{emtskl}). Plugging this new $\phi_{\alpha \beta}$ into that equation results for the terms with $p^\kappa$ and $S^{\kappa \lambda}$
\be p^\kappa v^\lambda \phi_{\kappa \lambda} + S^{\kappa \lambda}v^\mu \nabla_\kappa \phi_{\lambda \mu} = p^\kappa v^\lambda \nabla_{(\kappa} \phi_{\lambda)} + S^{\kappa \lambda} v^\mu \nabla_\kappa \nabla_{(\lambda} \phi_{\mu)}\ee
\be= \df{1}{2} p^\kappa v^\lambda \bigg( \nabla_\kappa \phi_\lambda + \nabla_\lambda \phi_\kappa\bigg) + \df{1}{4}S^{\kappa \lambda} v^\mu \bigg( \nabla_\kappa \nabla_\lambda \phi_\mu + \nabla_\kappa \nabla_\mu \phi_\lambda - \nabla_\lambda \nabla_\kappa \phi_\mu - \nabla_\lambda \nabla_\mu \phi_\kappa \bigg).\ee
This can written as
\be p^\kappa \df{\delta}{\delta s}A_\kappa + \df{1}{2} p^\kappa v^\lambda B_{\kappa
 \lambda} + \df{1}{4} S^{\kappa \lambda} v^\mu \bigg[ [\nabla_ \kappa , \nabla_
  \lambda] \phi_ \mu - [\nabla_ \mu ,\nabla_ \kappa] \phi_ \lambda + [\nabla_
   \mu , \nabla_ \lambda]\phi_ \kappa \bigg] + \df{1}{2} S^{\kappa \lambda}
    \df{\delta}{\delta s} B_{\kappa \lambda},\ee
where $A_\kappa$ and $B_{\kappa \lambda}$ are as defined in (\ref{yeniphiini}). Using the definition of Riemann tensor and the Bianchi identity, this can be put in the form
\be p^\kappa \df{\delta}{\delta s}A_\kappa + p^\kappa v^\lambda B_{\kappa \lambda} + \df{1}{2}S^{\kappa \lambda} \df{\delta}{\delta s}B_{\kappa \lambda} - \df{1}{2} {R_{\kappa \lambda \mu}}^\nu S^{\kappa \lambda} v^\mu A_\nu.\ee
For the rest of the terms in (\ref{emtskl}), remember that we have, by (\ref{yenitheta}),
\be \Phi_{\kappa \lambda} + {G^\mu}_{\kappa \lambda} \Lambda_\mu = \nabla_{\ast ( \kappa} \Phi_{\lambda)} - \nabla_{\ast (\kappa} \Theta_{\lambda)} + {H^\alpha}_\kappa {H^\beta}_\lambda \nabla_{(\alpha} \xi_{\beta)}.\ee
When plugged into (\ref{emtskl}) in this form, the first two terms on the right hand side do not contribute to the integral by the discussion of (\ref{vanishcov}). Recall that the left hand side of (\ref{emtskl}), with the choice $\phi_{\alpha \beta} = \nabla_\alpha \phi_\beta$ was shown to vanish previously. Then, the resulting equation is \cite{dixonIV}
\be \int ds \bigg[ p^\kappa \df{\delta}{\delta s}A_\kappa + p^\kappa v^\lambda B_{\kappa \lambda} + \df{1}{2} S^{\kappa \lambda} \df{\delta}{\delta s} B_{\kappa \lambda} - \df{1}{2}{R_{\kappa \lambda \mu}}^\nu S^{\kappa \lambda} v^\mu A_\nu + \int \skl \mathfrak{L}_\xi G_{\kappa \lambda} DX\bigg] = 0.\ee
Integrating by parts and collecting the terms $A_\kappa$, $B_{\kappa \lambda}$, and using the antisymmetry of the latter
\be \int ds \bigg\{ A_\kappa \bigg[ \df{\delta}{\delta s}p^\kappa + \df{1}{2} {R_{\nu \lambda \mu}}^\kappa S^{\nu \lambda} v^\mu \bigg] + B_{\kappa \lambda}\bigg[ \df{\delta}{\delta s}S^{\kappa \lambda}  - p^{[\kappa} v^{\lambda]}\bigg] -\df{1}{2} \int \skl \mathfrak{L}_\xi G_{\kappa \lambda} DX  \bigg\} = 0 .\ee
Using (\ref{forcegr}) and (\ref{torkcm}), this can be written as
\be \int ds \bigg[ A_\kappa F^\kappa + \df{1}{2}B_{\kappa \lambda} L^{\kappa \lambda} - \df{1}{2} \int \skl \mathfrak{L}_\xi G_{\kappa \lambda} DX \bigg]= 0.\ee
This holds for all $\phi_\alpha$ if the expression in the brackets vanishes identically. Then
\be A_\kappa F^\kappa + \df{1}{2}B_{\kappa \lambda} = \df{1}{2} \int \skl \mathfrak{L}_\xi G_{\kappa \lambda} DX, \label{yeniforce}\ee
for arbitrary $A_\kappa$ and $B_{\kappa \lambda} = B_{[\kappa \lambda]}$, with $\xi^\alpha$ as given in  (\ref{xiyeni}).
Comparing this final result with (\ref{cons1}) it can be seen that this is the simplest generalization of it. $\mathfrak{L}_\xi G_{\kappa \lambda}$ in normal coordinates with pole at $z$ reduces to $\mathfrak{L}_\xi g_{\alpha \beta}$, which vanishes when $\xi_\alpha$ is a Killing vector field. This shows that force and torque are results of the asymmetries of the spacetime.

The next thing to do is to find the explicit forms of $F^\kappa$ and $S^{\kappa \lambda}$ from (\ref{yeniforce}). When acting on an arbitrary two point tensor ${t^{\kappa ...}}_{\lambda ...}(z,x)$, the Lie derivative should be considered as acting on both of the arguments \cite{dixonIV}. Let us consider the action of the Lie derivative on a two point tensor which has scalar character at $x$
\be \mathfrak{L}_\xi {t^{\kappa ...}}_{\lambda ...} = \xi^\mu \nabla_\mu {t^{\kappa ...}}_{\lambda ...} + \xi^\alpha \nabla_\alpha {t^{\kappa ...}}_{\lambda ...} - (\nabla_\mu \xi^\kappa) {t^{\mu ...}}_{\lambda ...} - ... + (\nabla_\lambda \xi^\mu) {t^{\kappa ...}}_{\mu ...} + ... .\ee
Here, the second term is the outcome of the Lie derivative acting on the argument $x$, whereas the others follow from its action on the argument $z$. Now, as discussed in Appendix \ref{sec:Vertical and Horizontal Covariant Derivatives}
, one can treat ${t^{\kappa ...}}_{\lambda ...}$ as a function of $z$ and $X^\kappa = -\sigma^\kappa$ instead of $z$ and $x$. So, by using (\ref{xiinitial}), (\ref{jacgensol}), (\ref{vcdr}) and (\ref{horcd}); this can be put into the form
\be \mathfrak{L}_\xi {t^{\kappa ...}}_{\lambda ...} = A^\mu \nabla_\mu {t^{\kappa ...}}_{\lambda ...}  +\bigg({K^\alpha}_\mu A^\mu + {H^\alpha}_\mu \sigma^\lambda {B^\kappa}_\lambda \bigg)\nabla_\alpha {t^{\kappa ...}}_{\lambda ...} - {t^{\mu ...}}_{\lambda ...}{B_\mu}^\kappa - ... +{t^{\kappa ...}}_{\mu ...}  {B_\lambda}^\mu + ...          .\ee
\be = A^\mu \bigg(\nabla_\mu {t^{\kappa ...}}_{\lambda ...}+ {K^\alpha}_\mu \nabla_\alpha {t^{\kappa ...}}_{\lambda ...} \bigg) + {H^\alpha}_\mu \sigma^\lambda {B^\mu}_\lambda \nabla_\alpha {t^{\kappa ...}}_{\lambda ...}- {t^{\mu ...}}_{\lambda ...}{B_\mu}^\kappa - ... +{t^{\kappa ...}}_{\mu ...}  {B_\lambda}^\mu + ...         .\ee
\be A^\mu \nabla_{\mu \ast}  \tkl -   X^\lambda {B^\mu}_\lambda \nabla_{\ast \mu} {t^{\kappa ...}}_{\lambda ...}   - {t^{\mu ...}}_{\lambda ...}{B_\mu}^\kappa - ... +{t^{\kappa ...}}_{\mu ...}  {B_\lambda}^\mu + ...                .\ee
Applying this to $G_{\kappa \lambda}$, we have
\be \mathfrak{L}_\xi G_{\kappa \lambda} = A^\mu \nabla_{\mu \ast} G_{\kappa \lambda} + B^{\mu \nu} X_{\mu} \nabla_{\ast \nu} G_{\kappa \lambda} - {B^\mu}_\lambda G_{\kappa \mu} - {B^{\mu}}_\kappa G_{\mu \lambda}       .\ee
To put this in a more useful form
\be \mathfrak{L}_\xi G_{\kappa \lambda} = A^\mu \nabla_{\mu \ast}G_{\kappa \lambda} + 2B^{\sigma \nu} X_{\sigma} G_{\kappa \lambda \mu} + 2 \nabla_{\ast ( \kappa}(G_{\lambda ) \nu} B^{\mu \nu} X_{\mu})        ,\label{useful}\ee
where
\be G_{\kappa \lambda \mu \equiv} \df{1}{2} \nabla_{\ast \{ \kappa}G_{\mu \lambda\}}. \label{uclug}\ee
This reduces the $\Gamma_{\kappa \mu \lambda}$ in normal coordinates with pole at $z$, as can be seen from (\ref{curly}) and (\ref{buyukg}) \cite{dixonIV}, and satisfies
\be G_{\kappa \lambda \mu} = G_{\mu \nu}{G^\nu}_{\kappa \lambda}.\ee
In order to see this, we can simply expand the terms

\be \mathfrak{L}_\xi G_{\kappa \lambda} = A^\mu \nabla_{\mu \ast}G_{\kappa \lambda} + B^{\mu \nu} X_\nu \bigg( \nabla_{\ast \kappa} G_{\mu \lambda} - \nabla_{\ast \mu} G_{\lambda \kappa} + \nabla_{\ast \lambda} G_{\kappa \mu}    \bigg) + \nabla_{\ast  \kappa}(G_{\lambda \nu} B^{\mu \nu} X_{\mu}) + \nabla_{\ast  \lambda} (G_{\kappa \nu} B^{\mu \nu} X_\mu),            \ee
\be = A^\mu \nabla_{\mu \ast} G_{\kappa \lambda}  - B^{\mu \nu} X_{\nu}\nabla_{\ast \mu} G_{\lambda \kappa} + G_{\lambda \nu} \nabla_{\ast \kappa} (B^{\mu \nu} X_{\mu}) +  G_{\kappa \nu} \nabla_{\ast \lambda} (B^{\mu \nu} X_{\mu}).\ee
After cancelations and the use of (\ref{vercovder}), this reduces to (\ref{useful}). When used in this form in (\ref{yeniforce}), the last term does not contribute in view of the discussion of (\ref{vanishcov}). Then we are left with
\be A^\kappa F_\kappa + \df{1}{2}B^{\kappa \lambda}L_{\kappa \lambda} = \df{1}{2} \int \hat{T}^{\mu \lambda} \bigg( A^\kappa \nabla_{\kappa \ast} G_{\mu \lambda} + 2 B_{\mu \nu} X^\nu G_{\kappa \lambda \mu}\bigg) DX.\ee
Equating the coefficients, we have
\be F_\kappa = \df{1}{2} \int \hat{T}^{\lambda \mu} \nabla_{\kappa \ast} G_{\lambda \mu}DX,\label{forcegrav} \ee
\be L_{\kappa \lambda} = 2 \int \hat{T}^{\mu \nu} G_{\mu \nu [\kappa} X_{\lambda ]} DX,\label{torkgrav}\ee
in view of the antisymmetry of $B_{\kappa \lambda}$ \cite{dixonIV}.
This finalizes our discussion of gravitational force and torque on an extended body in a General Relativistic theory. The force and torque are obtained in their final form. Note that these definitions do not include $v^\kappa$, the center of mass velocity, so they are purely dynamical.
The next step in the procedure is to separate external and self forces, but this discussion is beyond the scope of this thesis. For the rest of this chapter, we will be working on the results of these equations on a test body and neglect the self-field on the object. Following such a line clearly excludes many physically interesting cases, such as the motion of a planet around the sun, but results will still be useful. For further information and different aspects of self force in theories of gravity, see \cite{poisson, waldself, dixonIV, abrahamtez}.

\subsection{Relativistic Multipole Moments}
\label{subsec:Relativistic Multipole Moments}
In order to evaluate the force and torque acting on a test body in General Relativity, we will make use of multipole moments, as we did in the Newtonian theory.
We will assume that the external field varies slowly over the body in concern \cite{dixonIV}.

For $n \geq 2$, we will define $2^n$-pole moment tensor of $\skl$ at $z(s)$ to be \cite{dixonIV}
\be I^{\kappa_1 \kappa_2 ... \kappa_n \lambda \mu} (s) \equiv \int X^{\kappa_1}X^{\kappa_2} ... X^{\kappa_n} \hat{T}^{\lambda \mu}(z,X)DX.\label{relmulmom}\ee
Clearly, it has the following symmetry
\be I^{\kappa_1 \kappa_2 ... \kappa_n \lambda \mu} = I^{(\kappa_1 \kappa_2 ... \kappa_n)( \lambda \mu)} .\ee
By using (\ref{sklcons}), another symmetry property follows: Using (\ref{vercovder}), we have
\be \nabla_{\ast \nu}(X^{\kappa_1}...X^{\kappa_n} X^{\kappa_\lambda} \hat{T}^{\nu \mu}) = (n+1)X^{(\kappa_1 ...} X^{\kappa_n}\hat{T}^{\lambda)\mu}.\ee
Integrating this over $X$, using (\ref{relmulmom}) and the fact that $\skl$ has compact support, the left hand side (boundary term) vanishes, then
\be I^{(\kappa_1 ... \kappa_n \lambda )\mu} = 0.\label{isym}\ee
In the above definition of the multipole moments, we have set $n \geq 2$. One can make the same definition for $n \geq 0$, but even this is done, $\skl$ will only contain quadrupole or higher order information about the body. On the other hand, one can extend the definition of the energy momentum skeleton and introduce an ``Extended Energy Momentum Skeleton $\skl_{ext}$" \cite{dixonIV} such that (\ref{emtskl}) can be written in the form:
\be \int \emt \phi_{\alpha \beta} \keg d^4x = \int ds \int \skl_{ext} (\Phi_{\kappa \lambda} + {G^{\mu}}_{\kappa \lambda} \Lambda_\mu)DX,\ee
where
\be \skl_{ext}= \skl (z,X) + p^{(\kappa}v^{\lambda )} \delta(X) - S^{\mu ( \kappa}v^{\lambda )} \nabla_{\ast \mu} \delta(X). \label{extskl}\ee
Here, $\delta(X)$ is the Dirac delta function on the tangent space of $z$. After the integration over $X$, this will give (\ref{emtskl}). If we had continued in this manner, we find that the multipole moments of $\skl_{ext}$ for $n= 0, 1$ are nonzero, and it can be seen in this formulation that momentum and angular momentum are the monopole and dipole moments of the energy momentum tensor. Following either line does not change anything in the following calculations, but an important point to remember that in Dixon's formalism, gravitational force and torque result from the quadrupole and higher order effects.

The moments (\ref{relmulmom}) contain all the information about $\skl$, and the latter can be obtained from the former by observing that, by (\ref{relmulmom}) \cite{dixonIV}
\be \int exp \bigg(i k_\kappa X^\kappa \bigg) \hat{T}^{\lambda \mu}DX = \sum_{n=2}^{\infty} \df{i^n}{n!} k_{\kappa_1} ... k_{\kappa_n} I^{\kappa_1 ... \kappa_n \lambda \mu}.\label{invfour} \ee
Where, the summation starts from $n=2$, since the lower moments vanish identically. $\skl$ can be obtained from this point by taking the inverse Fourier transform \cite{gelfand}. Note that taking the contraction of  (\ref{invfour}) with $k^\lambda$, and using (\ref{vercovder}) and (\ref{isym}), gives (\ref{sklcons}).
In order to obtain (\ref{jdekul}), we will introduce a new set of moments, $J^{\kappa_1 ... \kappa_2 \lambda \mu \nu \rho}$, which are related to $I$'s in the same fashion that $t'$'s are related to $t$'s in Newtonian mechanics (\ref{tprime}), and have symmetry properties analogous to (\ref{tprime1}),(\ref{tprime2}) and (\ref{tprime3}). It can be seen by using the definition of $J$'s in terms of $I$'s in (\ref{relmulmom}) that, using this new moments, (\ref{jdekul}) can be written in the form
\be n_{\kappa_1}J^{\kappa_1 ... \kappa_n \lambda \mu \nu \rho} = \quad  0, \quad \quad  n\geq 1.\ee
The main reason for introducing the $J$'s is the simplicity of such orthogonality properties. Besides, they have a more direct physical meaning than the $I$'s \cite{dixonIV}.

\subsection{Explicit Evaluation of Force and Torque}
\label{subsec:Explicit Evaluation of Force and Torque}
Assuming that the gravitational field varies slowly over the body, we can take the Taylor series expansion of (\ref{forcegrav}) and (\ref{torkgrav}) about $X=0$ \cite{dixonIV, dixonIII}. For the force expression, we have
\be  \nabla_{\kappa \ast} G_{\lambda \mu} = \sum_{n=0}^\infty \df{1}{n!} X^{\nu_1}...X^{\nu_n} \bigg[ \lim_{X \rightarrow 0} \bigg( \nabla_{\ast \nu_1 ... \nu_n } \nabla_{\kappa \ast} G_{\lambda \mu} \bigg)\bigg].\ee
Using this in (\ref{forcegrav}) with the use of (\ref{relmulmom}), results in
\be F_\kappa = \df{1}{2} \sum_{n=2}^\infty \df{1}{n!} I^{\nu_1 ... \nu_n \lambda \mu}\bigg[ \lim_{X \rightarrow 0} \bigg( \nabla_{\ast \nu_1 ... \nu_n } \nabla_{\kappa \ast} G_{\lambda \mu} \bigg)\bigg].\ee
Similarly, for the expansion of torque, we need
\be G_{\kappa \lambda \mu} = \sum_{n=0}^\infty \df{1}{n!} X^{\nu_1} ...  X^{\nu_n} \bigg[ \lim_{X \rightarrow 0} \bigg( \nabla_{\ast \nu_1 ... \nu_n } G_{\kappa \lambda \mu} \bigg)\bigg]. \ee
Using in (\ref{torkgrav}), again with (\ref{relmulmom})
\be L_{\kappa \lambda} = 2 \sum_{n = 1}^\infty \lim_{X \rightarrow 0} \bigg( \nabla_{\ast \nu_1 ... \nu_n } G_{\mu \nu [ \kappa} \bigg) {I_{\lambda ]}}^{\nu_1 ... \nu_n \mu \nu}. \label{oeh}\ee

By the discussion following (\ref{buyukg}), the limiting terms in the above expression reduces to ``extension of the metric tensor", which is discussed in Appendix \ref{sec:Tensor Extension}. Following their definition, it can be seen that the $\nabla_{\ast \kappa}$ and $\nabla_{\kappa \ast}$ commute \cite{dixonIV}, and as $X \rightarrow 0$,  $\nabla_{\kappa \ast} \rightarrow \nabla_\kappa $. So, we have
\be \lim_{X \rightarrow 0} \bigg( \nabla_{\ast \nu_1 ... \nu_n } \nabla_{\kappa \ast} G_{\lambda \mu} \bigg) = \nabla_\kappa \lim_{X \rightarrow 0} \bigg( \nabla_{\ast \nu_1 ... \nu_n }  G_{\lambda \mu} \bigg) = \nabla_\kappa g_{\lambda \mu, \nu_1 ... \nu_n},\ee
where $g_{\lambda \mu, \nu_1 ... \nu_n}$ denotes the $n^{th}$ extension of $g_{\lambda \mu}.$
Then, similarly for the expression in (\ref{oeh}), using (\ref{uclug})
\be \lim_{X \rightarrow 0} \bigg( \nabla_{\ast \nu_1 ... \nu_n } G_{\kappa \lambda \mu} \bigg) = \df{1}{2} g_{\{\kappa \lambda, \nu \}\nu_1 ... \nu_n}.\ee
When these results are plugged into (\ref{forcegr}) and (\ref{torkcm}), we have \cite{dixonIV, dixonIII}
\be \df{\delta}{\delta s}p_\kappa = \df{1}{2} v^\lambda S^\mu \nu R_{\kappa \lambda \mu \nu} + \df{1}{2} \sum_{n=2}^\infty \df{1}{n!} I^{\nu_1 ... \nu_n \lambda \mu} \nabla_\kappa g_{\lambda \mu, \nu_1 ... \nu_n},\ee
\be \df{\delta }{\delta s} S^{\kappa \lambda} = 2p^{[ \kappa} v^{\lambda ]} + \sum_{n=1}^\infty \df{1}{n!} g^{\sigma [ \kappa} I^{\lambda] \nu_1 ... \nu_n \mu \nu} g_{\{\sigma \mu , \nu\}\nu_1 ... \nu_n}.\ee
This series can be truncated at the required order of accuracy. If we consider the first nontrivial contributions to the gravitational force and torque, by using (\ref{gext2}), the result can be written in terms of $J$'s instead of $I$'s as \cite{dixonIV, dixonI,dixonIII}
\be \df{\delta}{\delta s}p_\kappa = \df{1}{2} v^\lambda S^{\mu \nu} R_{\kappa \lambda \mu \nu} + \df{1}{6}J^{\lambda \mu \nu \rho} \nabla_\kappa R_{\lambda \mu \nu \rho},\ee
\be \df{\delta }{\delta s} S^{\kappa \lambda} = 2p^{[ \kappa} v^{\lambda ]} - \df{4}{3} {R^{[\kappa}}_{\mu \nu \rho} J^{\lambda ] \mu \nu \rho} .\ee

% CHAPTER 1
\section{CONCLUSIONS}
\label{sec:Conclusions}

In this thesis, I have reviewed the problem of motion of extended objects in Newtonian and relativistic mechanics. This work mainly depends on the series of papers by W. G. Dixon published from 1970 to 1979 \cite{dixonIV, dixonI, dixonIII, dixonV, dixonII}. Even though I have discussed the problem in the presence of gravitational field, the theory can be extended to include electromagnetism as well. Dealing with the charge-current vector is actually a simpler problem when compared to the stress-energy tensor. Details of such a construction can be found in \cite{dixonII}.

After discussing the problem in Newtonian mechanics in Chapter \ref{sec:Newtonian Mechanics}, I have gone over the construction of momentum, angular momentum, mass center in a relativistic theory, and discussed the reasoning behind their christening as such by extracting the desired properties for these objects. The corresponding quantities in a maximally symmetric spacetime were calculated and shown to be meaningful in retrospect.

In the following sections, I have made a detailed analysis of the energy momentum skeleton, which completely describes the system just as good as the energy-momentum tensor does. Using these, I have also constructed the equations of motion for an extended body in a relativistic theory. It was shown that, as in the Newtonian theory, it is possible to define the torque and the force acting on the body, in a series involving the dynamical field. Note that in the sections \ref{subsec:Relativistic Multipole Moments} and \label{subsec:Explicit Evaluation of Force and Torque} we have also discussed the problem for a test particle ignoring the self-field, which greatly simplified the problem.

The problem of self force demands further discussion, but it is clearly a difficult one. If the self force on a body is involved in the calculations, it may not be possible to truncate the series expansion after a finite number of terms. In Chapter \ref{sec:General Relativity}, while making a multipole expansion of the external field acting on a test particle, we noted that taking a finite number of terms might be legitimate, if the field varies slowly over the body. This does not necessarily have to be the case for self-force. One might try to define the body such that the self field does not vary greatly over the body, but such a system may not have a physical counterpart in Nature.

The final equations of motions found in the Newtonian theory was shown to be indeterminate, since there were no constraints in the time evolution of the multipole moments. This problem was overcome by introducing the concept of rigidity, which resulted in determinate equations of motion. The same situation was also faced in the relativistic theory. In such a theory, there is no fully accepted concept of rigidity. A natural suggestion for such an object, resulting from Dixon's theory, is to require the components of the moments to be constant with respect to a comoving orthonormal frame defined along the central worldline \cite{ehlers}.

For some recent applications of Dixon's theory in different problems, one can check \cite{tod, singh, singh1, singh2, tod1, tucker, bini, semerak} for pole-dipole approximation, and \cite{bini1, bini2, bini3, abra} for quadrupole approximation.

\appendix
\section{WORLD FUNCTION}
\label{sec:world function}

Throughout this thesis, Synge's world function \cite{synge} have been used. It is a single valued function of each pair of points $(x_1, x_2)$, if there exists a unique geodesic connecting them. It is defined to be one half the geodesic distance between the points. World function and its properties in flat space will be discussed before writing down the definition in a general spacetime.
In flat space, following the above verbal definition, world function can be written as

$$\sigma(x,z) = \df{1}{2}(x-z)^\alpha(x-z)_{\alpha}, $$

since the geodesics in flat spacetime are straight lines.

Checking the first derivatives of this, we have:

\be \df{\partial \sigma(x,z) }{\partial x^\beta} = (x-z)_\beta   \quad, \quad \quad \df{\partial \sigma(x,z)}{\partial z^\beta} = -(x-z)_\beta, \label{skf} \ee

which are the separation vectors between points, one with a negative sign. Second derivatives obviously give Kronecker delta functions.

In order to generalize this to a curved spacetime, let $x(u)$ be the parametric form of a geodesic connecting $x_1 \equiv x(u_1)$ and $x_2 \equiv x(u_2)$ with $u$ an affine parameter along it. Then world function is defined as \cite{poisson, dixonI, synge}

\be \sigma(x_1, x_2) \equiv \df{1}{2}(u_2 - u_1) \int_{u_1}^{u_2} g_{\alpha \beta}(x(u)) \dot{x}^{\alpha}  \dot{x}^{\beta}  du,    \ee
with  $\dot{x}^{\alpha} \equiv \df{dx^\alpha}{du} \label{wf}.$

Note that using the geodesic equation, the integrand is constant on the geodesic, so this can be integrated to give $\df{\pm1}{2}s^2$, where $s$ is the length of the geodesic, and the sign depends on whether the geodesic is timelike/spacelike.

In order to evaluate the derivatives of the world function, we can check how $\sigma$ changes when one of the endpoints is moved \cite{poisson}.

Let us first vary $x_1$ as $x_1 \rightarrow x_1 + \delta x_1$ while keeping $x_2$ fixed. Assume that the new endpoint is connected to $x_2$ by a geodesic which is parametrized such that the affine parameter along it runs from $u_1$ to $u_2$. Then,

\be \delta \sigma (u_2 - u_1) = \df{1}{2}(u_2 - u_1)\int_{u_1}^{u_2} \bigg[ 2g_{\alpha \beta} \ua^\alpha \delta \ua^{\beta}   + g_{\alpha \beta, \gamma} \ua^\alpha \ua^\beta \delta x^\gamma    \bigg] du    .      \ee
Integrate the first term by parts
\be \delta \sigma (u_2 - u_1)[g_{\alpha \beta} \ua^\alpha \delta x^{\beta}]_{u_1}^{u_2} - (u_2 - u_1)   \int_{u_1}^{u_2} \bigg[g_{\alpha \beta} \ddot{x}^{\alpha}\delta x^\beta -\dfrac{1}{2}g_{\alpha \beta, \gamma} \ua^\alpha \ua^\beta \delta x^\gamma  + g_{\alpha \beta , \gamma}\ua^{\alpha}\ua^\gamma \delta x^{\beta}     \bigg]du.     \ee
The last two terms in the integral can be rearranged to give $\Gamma_{\gamma \alpha \beta}= \df{1}{2}(g_{\alpha \gamma, \beta} + g_{\beta \gamma, \alpha} - g_{\alpha \beta, \gamma})$
\be  = (u_2 - u_1)[g_{\alpha \beta} \ua^\alpha \delta x^{\beta}]_{u_1}^{u_2} - (u_2 - u_1)   \int_{u_1}^{u_2} \bigg[g_{\alpha \gamma} \ddot{x}^{\alpha} +  \Gamma_{\gamma \alpha \beta} \ua^{\alpha} \ua^{\beta}       \bigg]\delta x^\gamma du. \ee
So, the last term vanishes upon use of the geodesic equation. Evaluating the boundary term, keeping in mind that $x_2$ was kept fixed in the variation, we have

$$ \delta \sigma = -(u_2 - u_1) g_{\alpha \beta} \ua^{\alpha} \delta x^\beta \bigg|_{u_1}  ,   $$
\be \df{\partial \sigma}{\partial {x_1}^\beta} = -(u_2 - u_1) g_{\alpha \beta} \ua^{\alpha} . \label{wfz}  \ee
This is obviously a vector at $x_1$ and a scalar at $x_2$. Essentially the same calculation for the variation of $x_2$, keeping $x_1$ fixed, gives
\be \df{\partial \sigma}{\partial {x_2}^\beta} = (u_2 - u_1) g_{\alpha \beta} \ua^{\alpha} .   \label{wfx}\ee
The following shorthand notation for the covariant derivatives of the world function will be used from now on
$$\nabla_\alpha \nabla_\kappa \sigma = \sigma_{\kappa \alpha}.$$
For the notation, see Chapter \ref{sec:notation}.

So, (\ref{wfz}) can be written as \cite{dixonI, synge}
\be \sigma^\kappa = -u \dot{x}^\kappa \label{wfzz},\ee and
(\ref{wfx} as
\be \sigma^\alpha = u \dot{x}^\alpha \label{wfxx},\ee
where $u = u_2 - u_1$.
Note once again that after direct integration of (\ref{wf}) we have,
\be\sigma(x_1, x_2) = \df{1}{2} (u_2 - u_1)^2 g_{\alpha \beta}(x_1)\ua^{\alpha}(u_1)\ua^{\beta}(u_1) = \df{1}{2} (u_2 - u_1)^2 g_{\alpha \beta}(x_2)\ua^{\alpha}(u_2)\ua^{\beta}(u_2).\ee
So
\be 2\sigma = \sigma_\kappa \sigma^\kappa = \sigma_\alpha \sigma^\alpha \label{wfs},\ee
where we have used the mentioned index notation. So, $\sigma^\kappa$ $(\sigma_\alpha)$ is a vector tangent to the geodesic with length equal to the that of the geodesic. Also, taking the derivative of (\ref{wfz}) shows that
\be \sigma^\alpha = {\sigma^\alpha}_\beta \sigma^\beta\quad,\quad\quad \sigma^\kappa = {\sigma^\kappa}_\alpha \sigma^\alpha\quad,\quad\quad \sigma^\kappa = {\sigma^\kappa}_\lambda \sigma^\lambda. \label{wfss}\ee

Coincidence limits $(x\rightarrow z)$ of various two point tensors can be calculated with the help of (\ref{wfss}), and by assuming that the resulting tensor is independent of the direction in which the limit is taken \cite{poisson}. By (\ref{wf}),(\ref{wfzz}) and (\ref{wfzz}), we see that
\be \lim_{x \rightarrow z} \sigma = 0, \quad \lim_{x \rightarrow z} \sigma^\kappa = \lim_{x \rightarrow z} \sigma^\alpha = 0.\ee
For the limits of higher derivatives of the world function, consider, for example, the first equation in (\ref{wfss})
\be \sigma_\alpha = {\sigma_\alpha}_\beta \sigma^\beta \rightarrow g_{\alpha \beta} \sigma^\beta = {\sigma_\alpha}_\beta \sigma^\beta . \ee
Now use (\ref{wfx}) to get
\be (g_{\alpha \beta} - \sigma_{\alpha \beta})\ua^{\alpha} = 0.\ee
Here, using our assumption (the resulting tensor is independent of the direction in which the limit is taken) results that, in the coincidence limit $(x \rightarrow z)$,  $(\sigma_{\alpha \beta} \rightarrow g_{\alpha \beta})$. Similar calculation can be made for higher derivatives of the world function.

For the coincidence limit of two point tensors including a tensor index at $z$, in it convenient to employ Synge's rule \cite{synge}, which we will state without proof. A simple proof can be found in \cite{poisson}. Let $A$ be any two point tensor, with indices suppressed, and $<>$ denote the coincidence limit $(x \rightarrow z)$. The result of Synge is that \cite{dixonII}
\be <\nabla_\kappa A> = \nabla_{\kappa}<A> - {\delta^\alpha}_{\kappa}<\nabla_\alpha A>. \label{syngesrule}\ee

As a final point, we will discuss the relation of the world function to the exponential map. Assume that $X$ is a vector in the tangent space at $z$. Then one can set
\be X^\lambda = -\sigma^\lambda (z,x), \quad\mbox{if}\quad x = Exp_z X\label{expon}.\ee
Ho the exponential map is defined so that it takes a vector in the tangent space of $z$ to another point in the manifold which lies at a distance equal to the length of the vector.
Also exponential map can be used to define Riemann normal coordinates about a point $z$, which is separately discussed in Appendix \ref{sec:Tensor Extension} \cite{carroll}.

\section{VERTICAL AND HORIZONTAL COVARIANT DERIVATIVES}
\label{sec:Vertical and Horizontal Covariant Derivatives}

Throughout this thesis, two point functions are frequently used. In this section, we will define some natural derivation operations for such tensors. Consider a general two point tensor ${t^\lambda}_\mu(z,x)$. Now, instead of using $z$ and $x$, treat this as a function of $z$ and $X$, which can be done by (\ref{expon}) ($X$ was defined to be an element of the tangent space at $z$). Now, our aim is to define, instead of the covariant derivatives $\nabla_\alpha$ $\nabla_\kappa$; which are, as our notation, covariant derivatives at points $x$ and $z$, respectively; introducing covariant differentiation with respect to $z^\kappa$ and $X^\kappa$. Note that in $\nabla_\alpha (\nabla_\kappa)$, $z (x)$ is kept fixed during the differentiation. This will be our main standing point in the following definitions. We would like to do the same with $z$ and $X$ \cite{dixonIV}.
\subsection{Vertical Covariant Derivative}
The first derivative operation we would like to define will keep $z$ fixed, while changing the radius vector $X$ \cite{dixonIV}
\be \nabla_{\ast \kappa} \equiv \df{\partial}{\partial X}.\label{vercovder}\ee
Since $\nabla_\alpha$ also keeps $z$ fixed, it can be related to the above differentiation by the usual chain rule. For a two point tensor ${t^\lambda}_\mu$, which, by our notation convention, is a scalar at point $x$, and a rank 2 tensor at $z$, this relation can be written as
\be \nabla_\alpha {t^\lambda}_\mu = \df{\partial}{\partial x^\alpha} {t^\lambda}_\mu = \df{\partial X^\kappa}{\partial x^\alpha} \df{\partial}{\partial X^\kappa} {t^\lambda}_\mu = - \sigma^\kappa_\alpha \nabla_{\ast \kappa} {t^\lambda}_\mu \label{vcdr},\ee
where in the last equation, we used (\ref{expon}) and our notation for the covariant derivatives of the world function. This equation shows the general relation of $\nabla_{\ast \kappa}$ and $\nabla_\alpha$. Obviously, this is valid for two point tensors of arbitrary rank, which have scalar character at point $x$.
\subsection{Horizontal Covariant Derivative}
As stated before, we would like to define a derivation, which varies $z(s)$ while keeping $X$ constant. But this is not possible, since $X$ is defined to be a vector at the point $z$. Since this cannot be done, the best that could be done is to try to define a differentiation operation such that $X$ is parallelly transported along $z$, as $z$ is varied \cite{dixonIV}.
Parallel transport of $X$ along $z(s)$ implies
\be \df{\delta }{\delta s} X^\nu = \dot{X}^\nu + {\Gamma^\nu}_{\kappa\rho} \dot{z}^\kappa X^\rho = 0 \label{hd1},\ee
while for the derivative of ${t^\lambda}_\mu $ along $z$, one has
\be \df{\delta}{\delta s} {t^\lambda}_\nu = \df{d}{ds} {t^\lambda}_\nu + {\Gamma^\lambda}_{\kappa\mu}\dot{X}^\kappa {t^\mu}_\nu - {\Gamma^\mu}_{\kappa \lambda} \dot{z}^\kappa {t^\lambda}_\mu \label{hd2}    \ee
by the usual definition of differentiation along a curve. Expand the first term on the right hand side
\be \df{d}{ds}{t^\lambda}_\mu = \df{d z^\kappa}{ds} \df{\partial}{\partial z^\kappa}{t^\lambda}_\mu + \dot{X}^\kappa \df{\partial}{\partial X^\kappa} {t^\lambda}_\mu. \ee
Using (\ref{hd1}), this can be written as
\be \df{d}{ds}{t^\lambda}_\mu = \df{d z^\kappa}{ds} \df{\partial}{\partial z^\kappa}{t^\lambda}_\mu -{\Gamma^\kappa}_{\nu \rho} \dot{z}^\nu X^\rho \df{\partial}{\partial X^\kappa} {t^\lambda}_\mu, \ee
or
\be \df{d}{ds}{t^\lambda}_\mu = \df{d z^\kappa}{ds}\bigg[ \df{\partial}{\partial z^\kappa}{t^\lambda}_\mu -{\Gamma^\nu}_{\kappa \rho}  X^\rho \df{\partial}{\partial X^\nu} {t^\lambda}_\mu\bigg].\ee
Using this in (\ref{hd2}) yields
\be  \df{\delta}{\delta s} {t^\lambda}_\mu = \df{dz^\kappa}{ds}\bigg[ \df{\partial}{\partial z^\kappa} - {\Gamma^\lambda}_{\kappa\nu}\dot{X}^\kappa {t^\nu}_\mu - {\Gamma^\lambda}_{\kappa \mu} \dot{z}^\kappa {t^\lambda}_\nu {\Gamma^\nu}_{\kappa \rho} X^\rho \df{\partial}{\partial X^\nu} {t^\lambda}_\mu
     \bigg]  \equiv  \dot{z}^\kappa \nabla_{\kappa \ast} {t^\lambda}_\mu \label{hdt}.  \ee

If ${t^\lambda}_\mu$ does not have any $X$ dependence, the term in parenthesis reduces to the standard definition of the covariant derivative (note that the last term automatically vanishes in that case.). As in the previous case, this definition can be extended to two point tensors of arbitrary rank with scalar character at $x$.
Since this operation does not actually keep $X$ fixed, we can't express $\nabla_{\ast \kappa}$ in terms of $\nabla_\kappa$ only, but it can be written in a combination of $\nabla_\kappa$ and $\nabla_\alpha$. Using the parallel transport of $X$ in this operation, rewrite (\ref{hd1}) in terms of the world function. By (\ref{expon}) \cite{dixonIV}
\be 0 = \df{\delta}{\delta s} X^\nu = \df{\delta}{\delta s} \sigma^\nu (x,z) = \dot{z}^\kappa {\sigma^\nu}_\kappa + \dot{x}^\alpha {\sigma^\nu}_\alpha.\ee
Then,
\be \dot{x}^\alpha = {\sigma^\nu}_\kappa (-{\sigma^\nu}_\alpha)^{-1}\dot{z}^\kappa.\ee
Using (\ref{hktanim}), this can be written as:
\be \dot{x}^\alpha = {K^\alpha}_\kappa \dot{z}^\kappa. \label{hd3}\ee
We have obtained all the preliminary results needed. Now, evaluating the change of a two point tensor field ${t^\lambda}_\mu$, treating it as a function of $x$ and $z$, we get
\be \df{\delta}{\delta s} {t^\lambda}_\mu = \dot{z}^\kappa \nabla_\kappa {t^\lambda}_\mu + \dot{x}^\alpha \nabla_\alpha {t^\lambda}_\mu.\ee
Using (\ref{hd3}) for the last term, and remembering (\ref{hdt}), we get
\be \nabla_{\kappa \ast} {t^\lambda}_\mu = \nabla_\kappa {t^\lambda}_\mu + {K^\alpha}_\kappa \nabla_\alpha {t^\lambda}_\mu. \label{horcd}\ee

\section{JACOBI EQUATION}
\label{sec:Jacobi Equation}

Consider a general spacetime, with $x(u,v)$ a one parameter family of geodesics along it, such that $u$ is an affine parameter along each geodesic and $v$ labels the geodesics. Choose the following basis vectors adapted to a coordinate system
$$U^\alpha \equiv \df{\partial x^{\alpha}}{\partial u},$$
which is the tangent vector to the geodesics, and
$$V^\alpha \equiv \df{\partial x^{\alpha}}{\partial v}, $$
which is the deviation vector. $V^\alpha$ satisfies the geodesic deviation equation for each fixed $v$ \cite{carroll}
\be \df{\delta^2 V^{\alpha}}{\delta u^2} + {R^{\alpha}}_{\beta \gamma \delta} V^\delta = 0.  \label{egd}      \ee
Now, in the definition of the world function, set $z \equiv x(u_0 = 0)$ and $x \equiv x(u_0 = u)$, Then by (\ref{wfz})
\be \sigma^{\kappa}(z(0,v), x(u,v)) = -u U^{\kappa}(0,v)\ee
Differentiating this with respect to $v$ \cite{dixonI}, keeping in mind that the differentiation acts on each argument separately,
one obtains
\be {\sigma^{\kappa}}_\lambda V^\lambda + {\sigma^{\kappa}}_\alpha V^\alpha = -u \df{\delta U^\kappa}{\delta v} .      \ee
Then, $V^\alpha$ can be found as
\be  V^\alpha = (- {\sigma^\alpha}_\lambda)^{-1} {\sigma^\lambda}_\kappa V^\kappa - u    ( {\sigma^\alpha}_\kappa)^{-1} \df{\delta V^\kappa}{\delta u}. \label{egds}\ee
since $\df{\delta U^\kappa}{\delta v} = \df{\delta V^\kappa}{\delta u}$, which can most easily be checked by choosing a locally inertial frame \cite{synge}.

Consider this equation with $v=0$, so that $x(u,0)$ represents the geodesic along which $V^\alpha$ satisfies the equation of geodesic deviation. So (\ref{egd}) actually determines $V^\alpha$, in terms of $V^\kappa$ and $\df{\delta V^\kappa}{\delta u}$.  So, with the knowledge of these initial values at $z$, one can obtain its value at $x$.

Note also that a Killing vector field $\xi^\alpha$ also satisfies (\ref{egd}) which can be shown from (\ref{kvk}); but also by the defining equation (\ref{kve}). Its initial velocity, $\df{\delta \xi_{\kappa}}{\delta u}$, can be written as
\be \df{\delta \xi_{\kappa}}{\delta u} = \ua^\lambda \nabla_\lambda \xi_{\kappa} = \ua^\lambda \nabla_{[\lambda} \xi_{\kappa ]}. \ee
So, the values of $\xi^\kappa$ and $\nabla_{[\lambda} \xi_{\kappa ]}$ at a point can be used to find $\xi_\alpha$ at another point.
The derivatives of the world function appearing as coefficients in (\ref{egds}) are called the Jacobi propagators, and fields satisfying (\ref{egd}) along the geodesics are called Jacobi fields \cite{de felice}. For the propagators, the following notation is used
\be {H^{\alpha}}_\kappa \equiv (- {\sigma^\kappa}_\alpha)^{-1}\quad, \quad \quad  {K^{\alpha}}_\kappa \equiv {H^{\alpha}}_\lambda {\sigma^{\lambda}}_\kappa. \label{hktanim}\ee
With this notation, solution of \ref{egd} (geodesic deviation equation or Jacobi equation) with known initial values can be written as
\be \xi_\alpha(x) = {K_\alpha}^\kappa \xi_\kappa(z) + {H_\alpha}^\kappa \sigma^\lambda \nabla_{[\kappa} \xi_{\lambda]}(z).\label{jacgensol}\ee
 In flat space, following the limits of the derivatives of the world function, they reduce to ${\delta^\alpha}_\kappa$. Moreover, the coincidence limits ($x\rightarrow z$ case) will also be needed, which can be found as
\be \lim_{x \to z} {H^{\alpha}}_\kappa = \lim_{x \to z} {K^{\alpha}}_\kappa = {\delta^\alpha}_\kappa. \ee

It can be checked that $\xi^\alpha$ (not necessarily a Killing field) satisfying the Jacobi equation (\ref{egds}) will leave $\sigma^\kappa$ invariant under Lie derivation \cite{abrahamjacobi}. Since $\sigma^\kappa$ is a two point tensor, derivation should act on each parameter separately
\be \mathfrak{L}_\xi \sigma^\kappa = \xi^\lambda {\sigma^\kappa}_\lambda  + \xi^\alpha {\sigma^\kappa}_\alpha - \sigma^\lambda \nabla_\lambda \xi^\kappa = 0\label{sigmalie},\ee
since the term on the right hand side is identical to (\ref{egds}).

\section{TENSOR EXTENSION}
\label{sec:Tensor Extension}

Assume, in a manifold $M$, we have a point $p$, and another point $q$ which are connected by a geodesic $x^\mu(s)$, with $s$ an affine parameter along the geodesic. Then, $x^\mu$ satisfies
\be \ddot{x}^\mu + {\Gamma^\mu}_{\nu \tau} \dot{x}^\nu \dot{x}^\tau = 0. \label{geodes} \ee
In order to find a formal solution of this equation, one can take successive derivatives \cite{veblen}
\be \df{d^3 x^\mu}{ds^3} + \bigg( \df{d}{ds} {\Gamma^\mu}_{\nu \tau} \bigg)\dot{x}^\nu \dot{x}^\tau + {\Gamma^\mu}_{\nu \tau} \df{d}{ds}\bigg( \dot{x}^\nu \dot{x}^\tau\bigg)=0,\ee
\be  \df{d^3 x^\mu}{ds^3} + \bigg( \partial_\kappa {\Gamma^\mu}_{\nu\tau} \bigg)
\dot{x}^\nu \dot{x}^\tau \dot{x}^\kappa + {\Gamma^\mu}_{\nu \tau} \bigg( \ddot{x}^\nu
\dot{x}^\tau + \dot{x}^\nu \ddot{x}^\tau \bigg)=0,\ee
\be \df{d^3 x^\mu}{ds^3} + \dot{x}^\nu \dot{x}^\tau \dot{x}^\kappa \bigg( \partial_\kappa {\Gamma^\mu}_{\nu \tau} - {\Gamma^\mu}_{\alpha \tau} {\Gamma^\alpha}_{\nu \kappa} - {\Gamma^\mu}_{\nu \alpha} {\Gamma^\alpha}_{\kappa \tau}\bigg)\label{4gamma1} = 0.\ee
To write this in a more convenient form, let
\be  \partial_\kappa {\Gamma^\mu}_{\nu \tau} - {\Gamma^\mu}_{\alpha \tau}
{\Gamma^\alpha}_{\nu \kappa} - {\Gamma^\mu}_{\nu \alpha} {\Gamma^\alpha}_{\kappa
\tau} \equiv {\Gamma^{' \mu}}_{\nu \tau\kappa},\ee
and set
\be {\Gamma^\mu}_{\nu \tau \kappa} \equiv \df{1}{2}C({\Gamma^{' \mu}}_{\nu \tau\kappa}).\ee
Here $C$ denotes the cyclic permutation of the free subscripts. Then (\ref{4gamma1}) can be written as
\be \df{d^3x^\mu}{ds^3} + \dot{x}^\kappa \dot{x}^\nu \dot{x}^\tau {\Gamma^\mu}_{\nu \tau \kappa} = 0 . \ee
This procedure can be repeated for higher derivatives of (\ref{geodes}), and the equations will be of similar form
\be \df{d^4x^\mu}{ds^4} + {\Gamma^\mu}_{\nu \tau \kappa \lambda} \dot{x}^\nu \dot{x}^\tau \dot{x}^\kappa \dot{x}^\lambda  = 0,\ee
and so on. Here all the newly defined $\Gamma$'s will be symmetric in their subscripts.
If the initial conditions on (\ref{geodes}) are given as
\be x^\mu(0) = a^\mu \quad,\quad\quad \dot{x}^\mu (0) = b^\mu, \ee
then the exact solution of (\ref{geodes}) can be written as \cite{veblen}
\be x^\mu = a^\mu + b^\mu s - \df{1}{2}\bigg( {\Gamma^\mu}_{\nu \tau}|_a \bigg) b^\nu b^\tau s^2 - ... .\label{curve}\ee
Then, one can say that the transformation
\be x^\mu - a^\mu = y^\mu -\df{1}{2}\bigg( {\Gamma^\mu}_{\nu \tau}|_a \bigg) y^\nu y^\tau - ..., \ee
transforms the curves (\ref{curve}) into
\be y^\mu \equiv b^\mu s.\label{normal1}\ee
Hence it is possible to find a coordinate system in which curves through the origin satisfy (\ref{normal1}). Such a coordinate system will be called a ``normal coordinate system". Let the connection coefficients in such a coordinate system be ${\Gamma^{\ast \mu}}_{\nu \tau}$. Then if $y^\mu = b^\mu s$ is plugged into
\be \df{d^2 y^\mu}{ds^2} + {\Gamma^{\ast \mu}}_{\nu \tau} \dot{y}^\nu \dot{y}^\tau = 0, \ee
one has
\be {\Gamma^{\ast \mu}}_{\nu \tau} a^\nu a^\tau = 0,\ee
for all $a$'s. Multiplying with $s$, one obtains
\be {\Gamma^{\ast \mu}}_{\nu \tau} y^\nu y^\tau = 0.\ee
Expanding $\Gamma^\ast$'s in a power series of $s$'s
\be 0 = {\Gamma^{\ast \mu}}_{\nu \tau} y^\nu y^\tau = ({\Gamma^{\ast \mu}}_{\nu \tau})\bigg |_0 a^\nu a^\tau s^2 + \bigg( \df{\partial{\Gamma^{\ast \mu}}_{\nu \tau}}{\partial y^\kappa}\bigg)\bigg |_0 a^\nu a^\tau a^\kappa s^3 + ... .\ee
This holds only if
\be ({\Gamma^{\ast \mu}}_{\nu \tau})\bigg |_0 = 0,\ee
\be \bigg( \partial_{(\kappa}{\Gamma^{\ast \mu}}_{\nu \tau)}\bigg)\bigg |0 = 0,\ee
or for a term of arbitrary order
\be \bigg( \partial_{(\alpha \beta... \gamma} {\Gamma^{\ast \mu}}_{\nu \tau)}\bigg)\bigg |_0 = 0.\label{normalt}\ee
Now, this discussion \cite{veblen} shows that at the origin of normal coordinates, $\Gamma$'s vanish so that the covariant derivatives simply reduce to partial derivatives.  This can be extended to any tensor. We define the ``$k^{th}$ extension" of a tensor of rank $m,n$ as the rank $m, n+k$ tensor, denoted by ${T^{\alpha_1 ... \alpha_m}}_{\beta_1 ... \beta_n , \gamma_1 ... \gamma_r}$, (where the indices after the comma denotes extension indices) whose components at any point $(q)$ in any coordinate system $(x)$ is given as \cite{veblenthomas, veblen}
\be \bigg( {T^{\alpha_1 ... \alpha_m}}_{\beta_1 ... \beta_n , \gamma_1 ... \gamma_r}\bigg) \bigg|_{q} = \bigg( \partial_{\gamma_1 ... \gamma_n } {T^{\ast \alpha_1 ... \alpha_m}}_{\beta_1 ... \beta_n , \gamma_1 ... \gamma_r}	 \bigg)\bigg|_0. \label{normaltanim}\ee
Here, as can be understood from the notation, the right hand side involves expressions defined in the normal coordinate system. It can be shown that the $k^{th}$ extension of a rank $m,n$ tensor is a rank $m,n+k$ tensor by checking the transformation properties \cite{veblen}. Note that the resultant tensor is symmetric in its indices defining extension.
From the definition, it is obvious that the $1^{st}$ extension of a tensor is its covariant derivative.
Extensions of the connection coefficients  (called normal tensor) will be important in the following lines because of their direct relation to the normal coordinates. The extensions of the connections are tensors since their transformation between two normal coordinates with the same origin obeys the tensor transformation laws. From (\ref{normalt}) and (\ref{normaltanim}), it can be seen that
\be {\Gamma^\mu}_{(\nu \tau, \alpha_1 ... \alpha_n)} = 0. \label{extensionsym}\ee
We will use the following notation for the extensions of the connection coefficients
\be {A^\mu}_{\nu \kappa \tau} \equiv {\Gamma^\mu}_{\nu \kappa, \tau},\ee
\be {A^\mu}_{\nu \kappa \tau \lambda} \equiv {\Gamma^\mu}_{\nu \kappa , \tau \lambda},\ee
and so on.
In this thesis, we are interested in the extensions of the metric tensor. By metric compatibility, we have
\be \nabla_\mu g_{\alpha \beta} = 0.\label{compa}\ee
Then, the first extension of the metric tensor - which is the covariant derivative - vanishes. In order to calculate the higher extensions of it, one needs the help of the Riemann tensor. So we will postpone the discussion of higher extensions of $g_{\mu \nu}$ for now.
Riemann tensor in normal coordinates can be written as
\be {R^{\ast\mu}}_{\nu \tau \kappa} = \partial_\tau {\Gamma^{\ast\mu}}_{\nu \kappa} - \partial_\kappa {\Gamma^{\ast\mu}}_{\nu \tau} + \Gamma^{\ast2} + \Gamma^{\ast2} .\label{riemannnormal}\ee
When evaluated at the origin, the last two terms vanish, and in terms of the normal tensor, one has
\be {R^{\mu}}_{\nu \tau \kappa} = {A^{\mu}}_{\nu \kappa \tau} -  {A^{\mu}}_{\nu \tau \kappa}.\label{riemannnor}\ee
This can be inverted, by using the symmetries of $A$'s to get
\be {A^{\mu}}_{\kappa \nu \tau} = \df{1}{3}\bigg( {R^\mu}_{\nu \kappa \tau} - {R^{\mu}}_{\kappa \nu \tau}\bigg).\ee
On the other hand, differentiating (\ref{riemannnormal}) and evaluating at the origin of the normal coordinates gives
\be \nabla_\eta {R^{\mu}}_{\nu \tau \kappa} = {A^{\mu}}_{\nu \kappa \tau \eta} -  {A^{\mu}}_{\nu \tau \kappa \eta}.\ee
From this, by using the symmetry of $A$'s with respect to the first two indices, extension indices and with the help of (\ref{extensionsym}), one can obtain an expression of ${A^\mu}_{\nu \kappa \tau \eta}$ in terms of the covariant derivatives of the Riemann tensor \cite{veblen}
\be {A^\mu}_{\nu \tau \kappa \eta} = \df{1}{6}\bigg( 5\nabla_\eta {R^\mu}_{ \nu \kappa \tau } + 4\nabla_\kappa {R^\mu}_{ \nu \tau \eta } + 3\nabla_\tau {R^\mu}_{ \nu \eta \kappa} + 2\nabla_\nu {R^\mu}_{ \tau \kappa \eta} + \nabla_\nu {R^\mu}_{\tau \kappa \eta}\bigg).\label{5a}\ee
The procedure of expressing $A$'s in terms of $R$'s can be done to any order.
Now that we have all the necessary tools, we can attack the problem of finding the extensions of $g_{\mu \nu}$. This can be obtained by differentiating (\ref{compa}) in normal coordinates and evaluating at the origin. For the second extension, we have:
\be \nabla_{\beta \gamma}g_{\mu \nu} = 0 = \partial_\beta \bigg(\partial_\gamma g_{\mu \nu} - {\Gamma^\kappa}_{\gamma \nu}g_{\mu \kappa} - {\Gamma^\kappa}_{\gamma \mu}g_{ \kappa \nu}\bigg) + ...\quad \quad.\ee
The rest of the terms are of no interest to us since we will calculate this at the origin of the normal coordinates in which $\Gamma$'s vanish. Then, for the $2^{nd}$ extension, in terms of the normal tensors, one has
\be g_{\mu \nu, \gamma \beta} = g_{\mu \kappa}{A^\kappa}_{\gamma \nu \beta} + g_{\kappa \nu}{A^{\kappa}}_{\gamma \mu \beta}.\ee
This, in turn, can be written in terms of the Riemann tensor with the help of (\ref{riemannnor})
\be g_{\mu \nu, \gamma \beta} = -\df{1}{3}\bigg(R_{\mu \gamma \beta \nu} + R_{\mu  \beta\gamma \nu}	\bigg). \label{gext2}\ee
For the second extension, one has to consider
\be \nabla_{\alpha \beta \gamma} g_{\mu \nu}= 0 = {\partial_{\alpha \beta}}(\partial_\gamma g_{\mu \nu} - {\Gamma^\kappa}_{\gamma \nu}g_{\mu \kappa} - {\Gamma^\kappa}_{\gamma \mu}g_{\nu \kappa}) + ... \quad \quad ,\ee
in normal coordinates. The terms that were not explicitly written are the ones that will vanish at the origin of the normal coordinates, either because of vanishing connections, or $\partial_\gamma g_{\mu \nu} = 0$. Then, one has
\be g_{\mu \nu, \alpha \beta \gamma} = -g_{\mu \kappa} {A^{\kappa}}_{\gamma \nu \beta \alpha} - g_{\kappa \nu} {A^\kappa}_{\gamma \mu \beta \alpha},\ee
which, using (\ref{5a}), can be written as
\be g_{\mu \nu, \alpha \beta \gamma} = -\nabla_{(\alpha}R_{|\mu| \beta \gamma) \nu}. \label{gext3}\ee
One can obtain the extension of the metric tensor to any order by following the similar steps for higher derivatives of (\ref{compa}), and a closed form for the higher extensions can be found in \cite{dixonIV}.

\begin{acknowledgments}
The author is thankful to his supervisor Assoc. Prof. Dr. Bayram Tekin for
everything, but especially for suggesting him to study this delightful
subject. The author would also like to thank Assoc. Prof. Dr. B.
\"{O}zg\"{u}r Sar{\i}o\u{g}lu and Dr. A. Isaiah Harte for their guideance and  useful
discussion, and Enderalp Yakaboylu for his great assitance on almost all
parts of the thesis.
The author was the receipent of T\"{U}B\.{I}TAK Scholarship B\.{I}DEB-2228 throughout his graduate education.
\end{acknowledgments}

\end{document}